\documentclass[a4paper,12pt]{article}

\usepackage{times,amsfonts,epsfig,psfrag}
\usepackage[latin1]{inputenc}
\usepackage{amsfonts}
\usepackage{amsmath}
\usepackage{amssymb}
\usepackage{amsthm}
\usepackage[american]{babel}
\usepackage[T1]{fontenc}
\usepackage{slashed}
\usepackage{dsfont}
\usepackage{wick}

\newcommand{\qbar} {\overline{q}}

\newcommand{\bra}[1]{\left\langle #1 \right|} 
\newcommand{\ket}[1]{\left| #1 \right\rangle}

\newcommand{\nn}{\nonumber}  
\newcommand{\ubar}{\overline{u}}
\newcommand{\dbar}{\overline{d}}

\newcommand {\qcond} {\left\langle \qbar q\right\rangle} 
\newcommand{\Da}{\mathcal{D \underline{\alpha}} }
\allowdisplaybreaks

\begin{document}

\begin{titlepage}\begin{scriptsize}\end{scriptsize}

\begin{center}
{\LARGE \bf
Light-cone sum rules for the $N\gamma\Delta$ transitions for real photons}
\vspace{1cm}

{\sc J. \: Rohrwild}
\\[0.5cm]
\vspace*{0.1cm}{\it
Institut f\"ur Theoretische Physik, Universit\"at Regensburg,
\\D-93040 Regensburg, Germany} 
\\[1.0cm]

\vspace{0.6cm}
\bigskip
\centerline{\large \em \today}
\bigskip
%\vfill
\vskip1.2cm
{\bf Abstract:\\[10pt]} \parbox[t]{\textwidth}{
We examine the radiative $\Delta \to \gamma N$ transition at the real photon point $Q^2=0$ using the framework of light-cone QCD sum rules. In particular, the sum rules for the transition form factors $G_M(0)$ and $R_{EM}$ are determined up to twist 4. The result for $G_M(0)$ agrees with experiment within $10 \%$ accuracy. The agreement for $R_{EM}$ is also reasonable.
In addition, we derive new light-cone sum rules for the magnetic moments of nucleons, with a complete account of twist-4 corrections based on a recent reanalysis of photon distribution amplitudes.
}
   \vfill
{\em  }
\end{center}
\end{titlepage}

\thispagestyle{empty}
{\tableofcontents}
\newpage
\setcounter{equation}{0}

\section{Introduction}
\addtocounter{page}{-1}

The wish to understand the constituents of atomic nuclei, the nucleons, has been the driving force for a great many experiments and theoretical models. In particular the radiative nucleon-$\Delta$-transition has been in the focus of attention since 1979, when it was shown that a deformation of the nucleon-$\Delta$-system can entail non-vanishing electromagnetic (E2) and Coulomb (C2) quadrupole amplitudes \cite{Glashow:1979gp}. This breaks a selection rule laid down previously, which was derived in the first non relativistic quark model discussing the $E2$ amplitude \cite{Becchi} and which allows only magnetic dipole amplitudes (M1) in the $\gamma^*N\to \Delta$ transition.

The fact that the measurement of the  electromagnetic properties of the transition can provide insights in the deviation of the nucleon or the $\Delta$ from spherical symmetry has resulted in numerous experiments covering a large range of accessible values for the photon virtuality $Q^2$. In the whole region up to $\sim 4\;{\rm GeV}^2$ the ratios $E2/M1$ and $C2/M1$ are found to be small and negative, especially $|E2/M1|$ is smaller than $5\%$ \cite{Frolov:1998pw, Joo:2001tw, Kunz:2003we,Ungaro:2006df, Sparveris:2006uk}. In the case of real photons, which is relevant for this work, the Coulomb quadrupole is known to vanish. Thus precision measurements are only available for the ratio $E2/M1$ \cite{Blanpied:2001ae, Beck:1999ge, Sandorfi:1998xr, Beck:1997ew}.

On the theoretical side various approaches have been suggested. In quark models with unbroken $SU(6)$-spin-flavour symmetry $E2/M1$ is predicted to be exactly zero, whereas a broken $SU(6)$ symmetry yields values ranging from $0$ to $-2\%$ \cite{Isgur:1981yz, Gershtein:1981zf, Gogilidze:1986tu}. Other models, among them Skyrme models and the large $N_c$ limit of QCD, also find the ratio to be small and negative \cite{Wirzba:1986sc, Jenkins:2002rj}. 
Given that the $\Delta$ decays almost entirely into a nucleon and a pion, it is not surprising, that chiral bag models tend to agree well with experimental data \cite{Bermuth:1988ms}. In recent years studies using chiral effective field theory have been quite popular and yield rather precise results \cite{Gail:2005gz, Pascalutsa:2005vq}, in addition to that lattice calculations also predict $E2/M1$ to be around $-3\%$ \cite{Alexandrou:2004xn}. Only recently a detailed review summarising various theoretical approaches to the nucleon-$\Delta$-transition has been published \cite{Pascalutsa:2006up}.

On the other hand, the attempts to understand the nucleon-$\Delta$-transition at the microscopic level i.e. in terms of the underlying quark-gluon structure have been less successful. In particular, the calculation of Ioffe and Smilga \cite{Ioffe:1983ju} in the framework of QCD sum rules \cite{SVZ} failed to produce acceptable results. A possible reason for this is that the background field technique developed in \cite{Ioffe:1983ju} (see also \cite{Balitsky:1983xk} for an equivalent approach) is only applicable for the case that the participating initial state and final state hadrons have equal masses. Technically, this restriction arises because the  contribution of interest can only be isolated as the double-pole contribution in the hadron momentum. This is the case for e.g. the calculation of nucleon magnetic moments which was the primary task of \cite{Ioffe:1983ju, Balitsky:1983xk}, but it is not a good approximation for the $N \to \Delta\gamma$ radiative transition.

The problem of calculating the transitions between hadrons of different mass is known for a long time and provided the main motivation for the development of an alternative approach \cite{Balitsky:1989ry,Chernyak:1990ag}, now known as light-cone sum rules (LCSRs). In this technique, an infinite series of the ``induced condensates'' (in the language of \cite{Ioffe:1983ju}) is resummed in a function that has the physical meaning of a photon distribution amplitude and describes the probability amplitude to find a quark and an antiquark in the real photon, with given momentum fractions and at small transverse separations. The operator product expansion in light-cone sum rules is organised in terms of distribution amplitudes (DAs) of increasing twist. The relevant photon distribution amplitudes were introduced in \cite{Balitsky:1989ry} and recently studied in more detail in \cite{Ball:2002ps}. This technique has been used numerously, see e.g. \cite{Ball:2003fq, Aliev:2003ba, Dorokhov:2006qm, Colangelo:2005hv} for recent applications of photon DAs.

In this work we calculate the form factors of the $\gamma p \to \Delta^{+}$ transition at $Q^2=0$ using the light-cone sum rule formalism. 
In \cite{Aliev:1999tq} the transition has been studied at $Q^2=0$ with a LCSR approach similar to the one we use in this work. We use, however, a more recent complete set of photon DAs up to twist 4, that also include 3-particle-DAs containing an additional gluon, which do influence the final sum rules. These were also used in \cite{Aliev:2004ju}, where radiative decays of decuplet baryons into octet baryons have been considered and in particular the nucleon-$\Delta$ transition was also calculated. A comparison with this work is given in Section 3.
  As already stressed in previous works \cite{Belyaev:1995ya, Peters}, it is important to choose the Lorentz basis in such a way that the unwanted contributions due to the non-vanishing overlap of states with spin $1/2$ and negative parity with the $\Delta$ interpolating field $\eta^{\mu}$ can be separated from those of spin $3/2$ states with positive parity.

 For the calculation we will use a technique based on the background field method, that was first used in \cite{Balitsky:1989ry} and \cite{Braun:1988qv} to calculate the radiative $\Sigma \to p\gamma$ transition and the nucleon magnetic moments. In this work we will also give an update on the LCSR results for the magnetic moments.
 Our results for $\gamma p \to \Delta^+$ can easily be conferred to $\gamma n \to \Delta^{0}$ by exchanging $e_u \leftrightarrow e_d$ in the final formulae.

The present analysis is also fuelled by the results of Refs.\cite{Belyaev:1995ya, Peters}, where nucleon-$\Delta$-tran\-sition form factors were calculated for virtual photons. In the both calculations that use different (local duality and LCSR, using nucleon distribution amplitudes \cite{Braun:2000kw}) techniques, the magnetic transition form factor comes out to be below the data for the momentum transfers below $2 \;\!{\rm GeV}^2$, and the reason for this discrepancy is not understood. In order to understand the origin of this problem it is imperative to have an alternative calculation for the low $Q^2$ region. Our calculation for $Q^2=0$ provides a step in this direction.

The presentation is organised as follows. In Section 2 we will consider the $N\to \gamma N$ transition to calculate nucleon magnetic moments. The next Section deals with the  nucleon-$\Delta$-transition. We will not give the details of the calculation, but instead focus on the choice of an appropriate Lorentz basis. The sum rules for the magnetic dipole form factor $G_M(0)$ and the ratio $E2/M1$ are discussed in Section 4.

\section{Magnetic moments of nucleons}
\label{Proton}
In this Section we will examine the nucleon magnetic moments. This is a classical problem that provides a test ground for many non-perturbative methods. In particular, the calculation of nucleon magnetic moments was the main objective behind the generalisation of QCD sum rules in background fields \cite{Ioffe:1983ju, Balitsky:1983xk}. The results are in good agreement with experiments. In Refs.\cite{Braun:1988qv, Aliev:2002ra} the magnetic moments were already calculated using LCSRs in conjunction with photon DAs. As a new element, our calculation will use, for the first time, the complete set of photon DAs\footnote{We will not take into account 4-particle-DAs which are not expected to give rise to numerically relevant contributions} and also updated non-perturbative parameters.

This simple example serves as an illustration how the calculation is carried out and allows us to test its accuracy and the dependence of the results on the various parameters. Especially, this procedure provides a check for the numerical values of those non-perturbative parameters that are still under discussion, e.g. the magnetic susceptibility of the quark condensate $\chi$.

\subsection{Definitions}

\begin{figure}
\begin{center}
\epsfig{figure=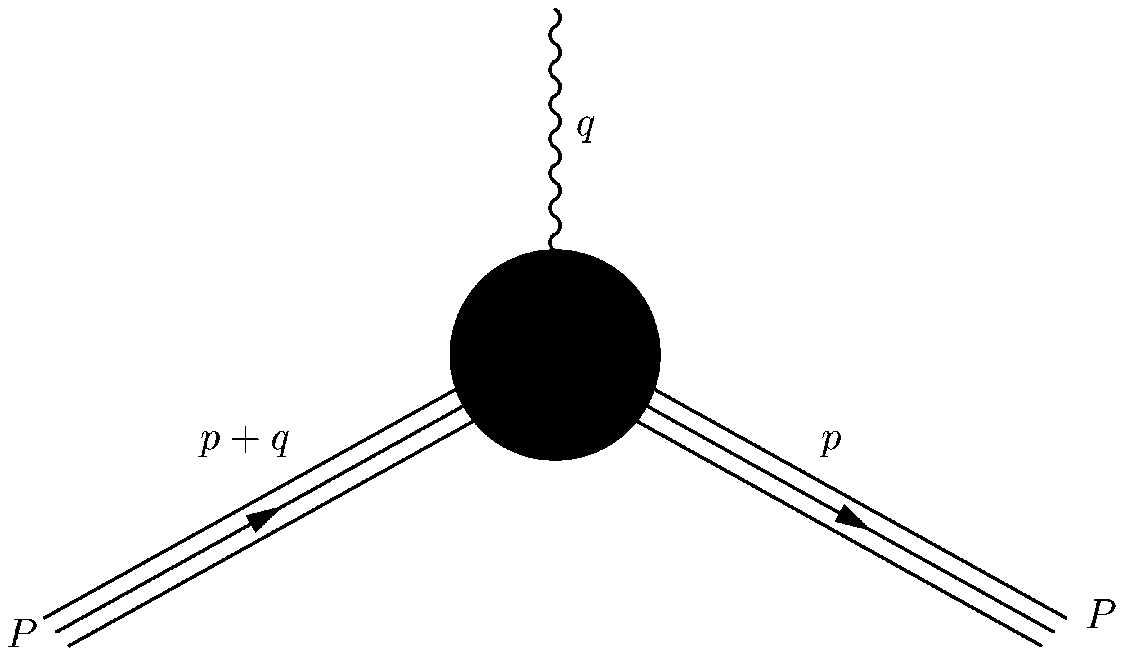,width=0.5\textwidth}
\end{center}
\caption{A proton with initial momentum $p+q$ emits a photon with momentum $q$. \label{PToP}}
\end{figure}
The transition matrix element  
\begin{align}
 	\label{Protonvertex}
	\bra{P(p,s)} j^{\mu}(0) \ket{P(p\hspace{-0.6pt}+\hspace{-0.6pt}q,s')} = P^{(s)}(p) \left[\gamma^{\mu} F_1(Q^2) -i \frac{1}{2m_p}\sigma^{\mu \nu} q_{\nu} F_2(Q^2) \right]\overline{P}^{(s')}(p\hspace{-0.6pt}+\hspace{-0.6pt}q) \mbox{ }
\end{align}
can conveniently be parametrised in terms of Dirac and Pauli form factors, $F_1$ and $F_2$.
Hereafter $P^{(s)}(p)$ is the proton spinor with momentum $p$ and spin $s$.
The magnetic moment of a nucleon can then be defined as
\begin{align}
\mu_N=F_1(0)+F_{2}(0) \mbox{ .}
\end{align}
This allows us to take only real photons into account.
As $F_{1}^{n}(0)=0$ and $F_{1}^{p}(0)=1$ are just the corresponding charges, it is only necessary to determine $F_{2}(0)$.

The process in Fig.\ref{PToP} can be described by the correlation function
\begin{align}
\label{ProtonKorrelator}
 \Pi^{\mu} \left(p,q \right) = i^2  \int\! d^4 x  \int \!d^4 y \;e^{i px + i qy} \bra{0} \mathcal{T} \lbrace \eta_{p} \! \left( x\right) j^{\mu}\left( y \right)\overline{\eta}_{p}\! \left( 0 \right) \rbrace  \ket{0} \mbox{ .}
\end{align}
Here 
\begin{align}
\label{EmCurrent}
j^{\mu}=e_d \dbar \gamma_{\mu} d + e_u \ubar \gamma_{\mu} u
\end {align}
is the electromagnetic current, with $e_d=-1/3$ and $e_u=2/3$ being the quark charges.
$e^{(\lambda)}_{\mu}$ is the four-polarisation vector of the emitted photon, $q \cdot e^{(\lambda)}=0$. The current 
\begin{align}
\label{IoffeCurrent}
\eta(x)=& \left(u^a(x)\mathcal{C}\gamma^{\lambda} u^{b}(x) \right) \gamma_5 \gamma_{\lambda }d^{c}(x)\varepsilon^{abc}
\end{align}
is the usual Ioffe current \cite{Ioffe:1981kw} for the proton. $\mathcal{C}$ is the charge conjugation matrix, $a,\;\!b,\;\!c$ are colour indices and $\varepsilon^{abc}$ is the three dimensional Levi-Civita symbol. Note that the current has fixed isospin $1/2$. The isospin relation between proton and neutron then assures that the formulae for the neutron magnetic moments can be obtained from those of the proton by exchanging $e_u \leftrightarrow e_d$. Therefore we will only consider the proton.
The coupling $\lambda_P$ of the Ioffe current Eq.\eqref{IoffeCurrent} to the proton is defined by
\begin{align}
\label{ProtonCoupling}
\bra{0} \eta(0)\ket{P(p,s)}=\frac{\lambda_{P}}{(2\pi)^2} P^{(s)}(p) \mbox{ .}
\end{align}

By introducing an electromagnetic background field of a plane wave
\begin{align}
\label{Fmunu}
F_{\mu \nu} = i\left(e^{(\lambda)}_{\nu} q_{\mu}-e^{(\lambda)}_{\mu} q_{\nu} \right) e^{iqx}=f_{\mu \nu} e^{iqx} \mbox{ ,}
\end{align}
it is possible to absorb the emitted photon into the background field.
We consider the following object
\begin{align}
 \label{Protoncorrelator}
 \Pi_{P}^{\mu} \left(p,q \right) e^{\left( \lambda \right)}_ {\mu}= i \int\! d^4 x   \;e^{i px} \bra{0} \mathcal{T} \lbrace \eta_{p} \! \left( x\right) \overline{\eta}_{p}\! \left( 0 \right) \rbrace  \ket{0}_F \mbox{ .}
\end{align}
Here the subscript $_{F}$ indicates that the v.e.v. has to be evaluated in the background field $F_{\mu \nu}$. Expanding the correlator in Eq.\eqref{Protoncorrelator} in powers of the background field and taking only the terms linear in $F_{\mu \nu}$ corresponds to the single photon emission, also described by Eq.\eqref{ProtonKorrelator}. We will refrain from a more detailed presentation of this expansion, but instead would like to refer to \cite{Novikov:1983gd} for a review on the background field method and to Chapter 2 in Ref. \cite{Ball:2002ps}, which is dedicated to the expansion of correlation functions in an electromagnetic background field.

\subsection{Expansion of the correlator}

Applying the usual strategy of QCD sum rules, we have to calculate the correlation function Eq.\eqref{Protoncorrelator} in two different regimes. 
If the in- and out-going particle states are close to proton mass-shell, i.e. $p^2\approx m_P^2$ and $(p+q)^2 \approx m_P^2$, the hadronic representation of Eq.\eqref{Protoncorrelator} will be dominated by the process $p \to p \gamma$. 

The contribution $T_P$ of the process $p \to p \gamma$ is then given by
\begin{align}
T_P(p,q)=e^{(\lambda)}_{\mu} \frac{\bra{0} \eta_P(0) \ket{P(p,s)} \bra{P(p,s)} j^{\mu}(0) \ket{P(p+q,s)} \bra{P(p+q,s)} \overline{\eta}(0) \ket{0}}{\left(m_P^2-p^2 \right) \left(m_P^2-(p+q)^2 \right)}
\end{align}
Using Eqs.\eqref{Protonvertex},\eqref{IoffeCurrent} and the spin summation formula for Dirac spinors
\begin{align}
\label{spinsum1}
\sum_{s} P^{(s)}(p) \overline{P}^{(s)}(p) = & \; \slashed{p}+m_{P}
\end{align}
this can be written as
\begin{multline}
\label{Pcontrib}
T_{P}^{\mu} \left(p,q \right) e^{\left( \lambda \right)}_ {\mu}=\frac{|\lambda_P|^2}{(2\pi)^4} \frac{\slashed{p}+m_p}{m_p^2-p_1^2} \left[\gamma^{\mu} F_1(Q^2) -i \frac{1}{2m_p} \sigma^{\mu \nu} q_{\nu} F_2(Q^2) \right] \frac{\slashed{p}+\slashed{q}+m_p}{m_p^2-p_2^2} e^{\left( \lambda \right)}_{\mu} \mbox{ ,}
\end{multline}
with $p_1=p$ and $p_2=p+q$.\\
The Lorentz structure
\begin{align}
	\slashed{p} \sigma^{\mu \nu} \slashed{p} q_{\nu} e_{\mu}^{(\lambda)} \nn
\end{align}
is free of contributions due to $F_1(0)$, already satisfies the Ward identity, as it is proportional to $f^{\mu \nu}$, and has the highest possible number of momenta $p$. Therefore it seems advisable to focus on structures containing an even number of $\gamma$-matrices. 

After reducing to the Dirac basis we will only consider the structure $p^{\alpha} p_{\beta} \sigma^{\beta \zeta} f_{\alpha \zeta}$, following \cite{Braun:1988qv}.
Then one gets for Eq.\eqref{Pcontrib} 
\begin{align}
\label{ProtonOnHadron}
T_{P}^{\mu} \left(p,q \right) e^{\left( \lambda \right)}_ {\mu}= -\left( \frac{|\lambda_p|^2}{(2\pi)^4 m_p(m_p^2-p_1^2)(m_p^2-p_2^2)}F_2(0) \right)p^{\alpha} p_{\beta} \sigma^{\beta \zeta} f_{\alpha \zeta}+\ldots \mbox{ .}
\end{align}
The dots represent terms of different Lorentz structure.

\begin{figure}[t]
\centering{\epsfig{figure=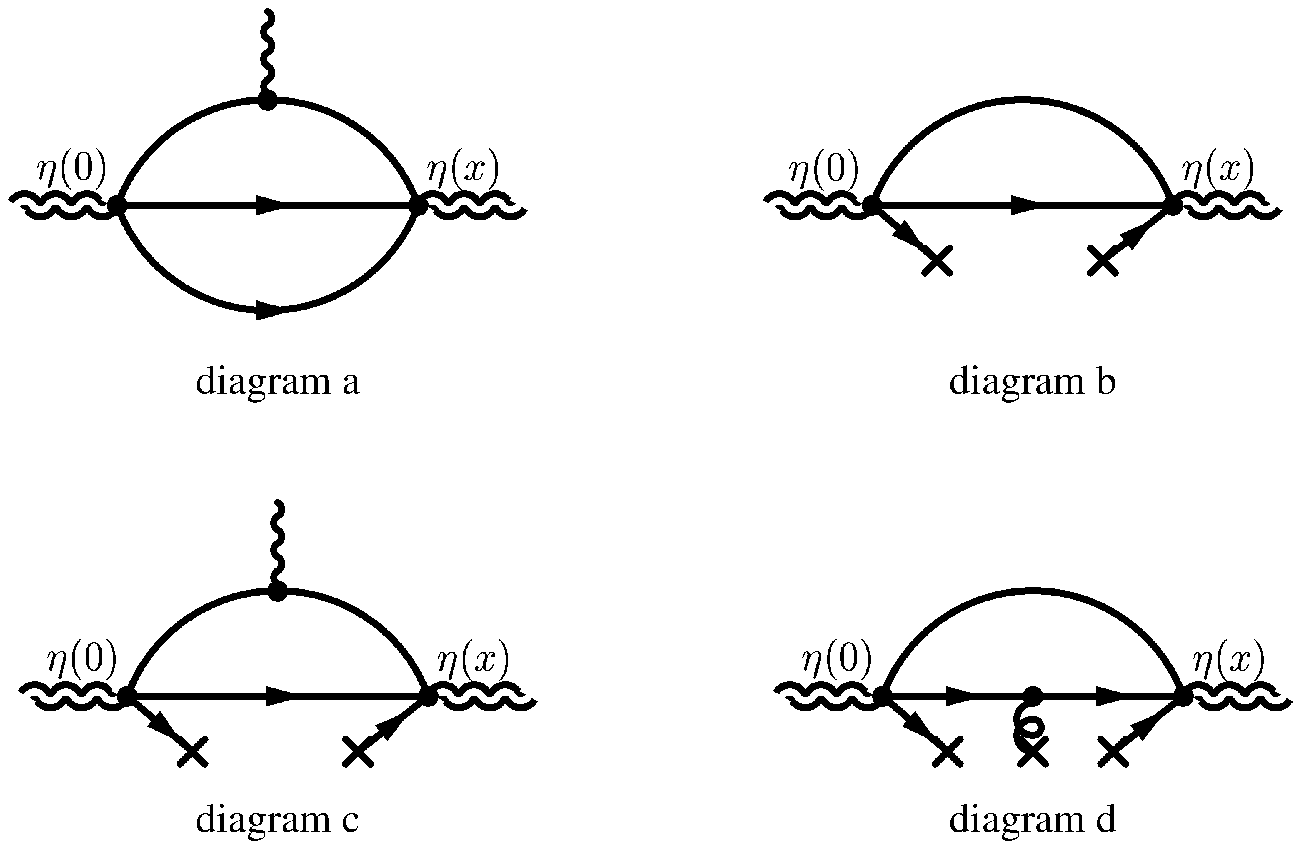,width=0.8\textwidth}}
\caption{Diagrams up to twist 4. The wiggled and the curled lines represent the coupling to the electromagnetic and gluonic background fields. The crosses denote  interactions with the vacuum. \label{FD}}
\end{figure}

In the Euclidean region, where $p_1^2 \ll 0$ and $p_2^2 \ll 0$, one can express the correlation function Eq.\eqref{ProtonKorrelator} in terms of photon distribution amplitudes of increasing twist. To this end we insert the expressions for the current $\eta$, Eq.\eqref{IoffeCurrent}, into the correlator \eqref{Protoncorrelator}. 
\begin{multline}
\Pi^{\mu \nu}(p,q) e^{\left( \lambda \right)}_ {\nu}= i\int \!\! d^4x\; e^{ipx+iqy} \bra{0}\mathcal{T} \left \lbrace  \left( u^a(x) \mathcal{C} \gamma^{\lambda} u^{b}(x) \right)  \gamma_5 \gamma_{\lambda} d^{c}(x) \cdot \right.
\\ 
\left.\cdot\; \dbar^{c'}\left( 0\right)\gamma^{\lambda'} \gamma_5 \left(\ubar^{a'}\left( 0 \right) \gamma_{\lambda'} \mathcal{C}\ubar^{b'}\left( 0 \right) \right)   \right \rbrace \ket{0}_F \varepsilon^{abc} \varepsilon^{a'b'c'}
\end{multline}
Using Wick's theorem the calculation is straightforward. One has, however, to pay attention to the fact that we are working with massless quarks in a simultaneous electromagnetic and gluonic background field. The electromagnetic field $F_{\mu \nu}$ is just a plane wave, whereas the gluonic field $G_{\mu \nu}=G_{\mu \nu}^A t^A$ due to the presence of gluons in the hadron is unknown. The quark propagator then adopts the following form\footnote{Note that our sign convention for the electric charge, $e=\sqrt{4 \pi \alpha_{em}}$, follows \cite{Ball:2002ps} and thus differs from \cite{Braun:1988qv}.} \cite{Balitsky:1987bk}:
\begin{align}
\label{propagator}
\wick{1}{<1{q}(x) >1{\overline{q} }}(0)=&\frac{i\slashed{x}}{2 \pi^2 x^4}\left[x,0 \right]-\frac{ig}{16 \pi^2 x^2}\int_0^1 \! du \; \left[x,ux \right]  \left \lbrace \ubar \slashed{x} \sigma_{\alpha \beta} + u  \sigma_{\alpha \beta} \slashed{x} \right \rbrace G^{\alpha \beta}(ux) \left[ux,0 \right]  \nn
\\
&-\frac{ie_q }{16 \pi^2 x^2} \int_0^1 \!du\; \left[x,ux \right] \left \lbrace \ubar \slashed{x} \sigma_{\alpha \beta}+u  \sigma_{\alpha \beta}\slashed{x} \right \rbrace F^{\alpha \beta}(ux)\left[ux,0 \right]+\ldots \mbox{  ,}
\end{align}
here we used the abbreviation
\begin{equation}
\label{pexp}
\left[ x,y \right]={\rm Pexp}\left \lbrace i \int_0^1 \! dt \;(x-y)_{\mu} \left[ e_q A^{\mu}(tx-\overline{t}y) + g B^{\mu}(tx-\overline{t}y) \right] \right \rbrace
\end{equation}
for the path-ordered exponent ($A_{\mu}$ is the electromagnetic and $B_{\mu}$ the gluon field) and $\overline{a}=1-a$, $\forall a \in [0,1]$. The dots represent terms that will give rise to contributions of twist 5 or higher. As we will only consider terms up to twist 4, these are not relevant here. Hence, there are only four diagrams that have to be taken into account, see Fig.\ref{FD}. 

 It turns out that the diagram in Fig.2a does not give rise to contributions with an even number of $\gamma$-matrices and can be neglected.

 After using Fierz identity to decompose $\bra{0} q_i(x) \qbar_j(0)\ket{0}_F$, one can insert the expression for the photon DAs (see Appendix \ref{A}) and perform the Fourier transformation. \\
We get for diagram b:
 \begin{align}
\label{diagramBProton}
	T^{P}_b(p,q) =& \left[ - \frac{ e_d \qcond } { 6 \pi^2  } \int_0^1 \! du \;\varphi(u) \ln \left(\frac{\mu^2}{-\ubar p_1^2-u p_2^2} \right)\right. 
\nn\\
	&\left. \phantom{\Big[}+\frac{e_d \qcond}{8 \pi^2} \int_0^1 \! du\; \frac{\mathds{A}(u)+\mathds{B}(u)}{-\ubar p_1^2-up_2^2} \right] p^{\alpha} p_{\beta} \sigma^{\beta \zeta} f_{\alpha \zeta} + \ldots \mbox{ ,}
\end{align}
here the dots represent terms that do not contribute to the structure $p^{\alpha} p_{\beta} \sigma^{\beta \zeta} f_{\alpha \zeta}$ or that are just polynomials in $p_1^2$ and $p_2^2$. These  will vanish after a subsequent double Borel transformation. 

The calculation of the remaining two diagrams is analogous and yields
\begin{align}
\label{diagramCProton}
T^{P}_{c}(p,q)=&\left[\frac{e_u \qcond}{2\pi^2} \int_0^1 \!\!du\;\int \Da 
\frac{S_{\gamma}(\underline{\alpha})}{-\overline{\alpha}_u p_1^2-\alpha_u p_2^2} \right.
\nn \\ 
& \phantom{\Big[}\left. -\frac{e_u}{2 \pi^2}\int_0^1\!\!du\;\int \Da \frac{(1-2u) T_{4}^{\gamma}(\underline{\alpha})}{{-\overline{\alpha}_up_1^2-\alpha_u p_2^2}} \right] p^{\alpha} p_{\beta} \sigma^{\beta \zeta} f_{\alpha \zeta} + \ldots \\ 
\label{diagramDProton}
T^{P}_{d}(p,q)=&\left[-\frac{e_d \qcond}{4 \pi^2} \int_0^1 \!\! du \; \int \!\! \Da \; \frac{\mathcal{S}(\underline{\alpha})+\widetilde{\mathcal{S}}(\underline{\alpha})}{-\overline{\alpha}_u p_1^2-\alpha_u p_2^2}  \right.
\nn \\
&\phantom{\Big[}\left. +\frac{e_d \qcond}{4 \pi^2} \int_0^1 \!\! du \; \int \!\! \Da \; (1-2u)\frac{T_2(\underline{\alpha}) - 2T_3(\underline{\alpha}) + T_4(\underline{\alpha})}{-\overline{\alpha}_u p_1^2-\alpha_u p_2^2} \right] 
\nn \\
&\phantom{ +\frac{e_d \qcond}{4 \pi^2} \int_0^1 \!\! du \; \int \!\! \Da \;} \times p^{\alpha} p_{\beta} \sigma^{\beta \zeta} f_{\alpha \zeta} + \ldots \mbox{ ,}
\end{align}
where $\int Da = \int_0^1 d \alpha_q \;\int_0^1 d\alpha_{\qbar}\;\int_0^1d \alpha_{g} \; \delta(1-\alpha_q-\alpha_{\qbar}-\alpha_{g})$. The functions $\varphi$, which is of twist 2 and $T_i$, $\mathcal{S}$, $\widetilde{\mathcal{S}}$, $\mathcal{S}_{\gamma}$, $T_{4}^{\gamma}$, $\mathds{A}$ and $\mathds{B}$, which have twist 4, are defined in Appendix\ref{A}.

\subsection{Borel transformation and continuum subtraction}

The sum rule for $F_2(0)$ can readily be obtained be equating the hadronic result, Eq.\eqref{ProtonOnHadron}, and the light-cone expansion,  Eqs.(\ref{diagramBProton}, \ref{diagramCProton}, \ref{diagramDProton}).
As usual, a Borel transformation and a subsequent continuum subtraction are necessary to suppress the effects of excited states and extract the $p \to p \gamma$ ratio. The two independent momenta $p_1$ and $p_2$ allow a double Borel transformation, that can be performed using the general formulae 
\begin{align}
\label{PBorel1}
\mathcal{B}_{M_1^2}\mathcal{B}_{M_2^2}\left \lbrace\frac{\Gamma(\alpha)}{\left(-\ubar p_1^2-u p_2^2 \right)^{\alpha}} \right \rbrace &= t^{2-\alpha} \delta\left(u-\frac{M_1^2}{M_1^2+M_2^2} \right)
\end{align}
and 
\begin{align}
\label{PBorel2}
\mathcal{B}_{M_1^2}\mathcal{B}_{M_2^2}\left \lbrace\frac{1}{\left(m^2_{1}-p_1^2 \right)\left(m^2_{2}-p_2^2 \right)} \right \rbrace &= e^{-m_{1}^2/M_1^2-m_2^2/M_2^2}\mbox{ .}
\end{align}
Here $M_i^2$ is the Borel parameter corresponding to $p_i^2$ and $t=\frac{M_1^2 M_2^2}{M_1^2+M_2^2}$.
The continuum subtraction can be accomplished by a simple set of substitution rules \cite{Belyaev:1994zk}:
\begin{eqnarray}
\label{SubstitionOne}
t^3 & \longrightarrow & t^3\left(1-e^{-S_0/t}\left(1+\frac{S_0}{t} +\frac{1}{2} \left(S_0/t \right)^2 \right) \right)\\
\label{SubstitionTwo}
t^2  & \longrightarrow & t^2 \left(1-e^{-S_0/t}\left(1+\frac{S_0}{t} \right) \right)\\
\label{Substitionthree}
t   &\longrightarrow & t\left(1-e^{-S_0/t}\right)
\end{eqnarray}
with $S_0$ being the continuum threshold.\\
For the process $p \to p \gamma$ the natural choices are
\begin{align}
\label{BorelratioProton}
	&M_1^2=M_2^2  
\end{align}
and the threshold $S_0$ coincides with the normal continuum threshold $s_P$ for the proton. 

Putting everything together the final sum rule for $F_2(0)$ takes the following form:
\begin{align}
\label{finalSR}
F_2(0)=& \frac{8 \pi^2  m_p \qcond}{|\lambda_P|^2} e^{m_p^2/t}\bigg[ \frac{e_d}{3} \varphi(1/2) t^2 \left(1-e^{-S_0/t}\left(1+\frac{S_0}{t} \right) \right) 
\nn \\
&+\left[-\frac{e_d}{4}\left( \mathds{A}(1/2)+\mathds{B}(1/2) \right) \right] t \left(1-e^{-S_0/t} \right)
\nn \\
&-\frac{e_d}{2}\int_0^{1/2} \!\! d\alpha_q \int_0^{1/2}\!\! d\alpha_{\qbar} \frac{1}{1-\alpha_q-\alpha_{\qbar}} 
\nn\\
&\qquad\quad\times \!\left[\mathcal{S}+\widetilde{\mathcal{S}}+\frac{2e_u}{e_d} \mathcal{S}_{\gamma}\right](\alpha_q,\alpha_{\qbar},1\!-\alpha_q\!-\alpha_{\qbar}) t \left(1-e^{-S_0/t} \right)
\nn \\
&+\frac{e_d}{2}\int_0^{1/2} \!\! d\alpha_q \int_0^{1/2}\!\! d\alpha_{\qbar} \frac{\alpha_q-\alpha_{\qbar}}{(1-\!\alpha_q-\!\alpha_{\qbar})^2}
\nn\\
&\qquad\quad \times \!\left[\! {T}_2-2{T}_3+{T}_4 -\frac{2e_u}{e_d}{T}^{\gamma}_4\! \right]\!(\alpha_q,\alpha_{\qbar},1\!-\alpha_q\!-\alpha_{\qbar}) t \left(1-e^{-S_0/t} \right) \bigg] \mbox{ .}
\end{align} 
The first term in Eq.\eqref{finalSR} gives the leading twist-2 contribution, which was first obtained in Ref.\cite{Braun:1988qv}. The remaining terms are new.

\subsection{Numerical Results}

 The asymptotic expression for the photon wave function $\varphi(u)$ is given by $\chi(\mu) 6 u (1-u)$ at a renormalisation scale $\mu^2=1\;\!{\rm{GeV}}^2$, where $\chi(\mu)$ is the so called magnetic susceptibility of the quark condenstate. It has been argued that the full DA does not differ much from the asymptotic expression \cite{Braun:1988qv} as sum rule calculations showed a small coefficient for the next-to-leading order term. Thus, henceforth, we will use $\varphi(1/2)=3/2 \chi(\mu)$.

 The value of $\chi$ is not very well known. The first detailed study using QCD sum rules yielded $\chi(1/2)=4.4 \;\!{\rm{GeV}}^{-2}$ \cite{Belyaev:1984ic,Balitsky:1985aq}, whereas a local duality approach \cite{Balitsky:1983xk} found  $\chi(1/2) \approx 3.3\;\!{\rm{GeV}}^{-2}$. The latest estimate for $\chi(\mu=1{\rm GeV})$ gives a value of $3.15 \;\!{\rm GeV}^{-2}$ \cite{Ball:2002ps} and, assuming asymptotic DAs, leads to $\varphi(1/2)\approx 4.73 {\rm GeV}^{-2}$.

  The higher-twist photon DAs are known to next-to-leading order in conformal spin. The corresponding expressions are collected in Appendix \ref{A}. 

\begin{figure}[t]
\begin{center}
\vspace{5mm}
\epsfig{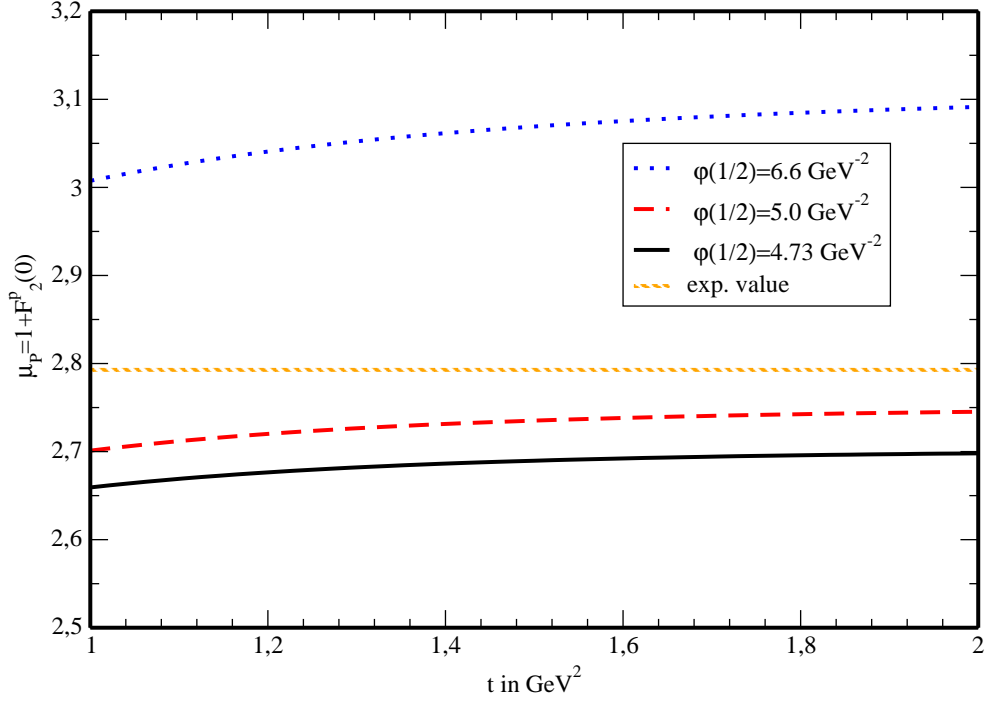} 
\end{center}
\caption{Magnetic moment of the proton from sum rule \eqref{finalSR} for different values of $\varphi(1/2)$. The hatched line represents the experimental value  \cite{PDG}. \label{ProtonGraph}}
\end{figure}

Furthermore, we need the numerical values for the coupling constant $|\lambda_{P}|^2$, the continuum threshold $s_{P}$ as well as the Borel window for $t$. The continuum threshold can be determined from the most fundamental sum rule for the nucleon, namely those for the coupling constant. We obtain $s_{P}\approx 2.25\;\!{\rm{GeV}}^2$, see \cite{Belyaev:1982sa}. It is advantageous to use the whole sum rule \cite{Braun:2006hz} instead of a fixed number for $|\lambda_{P}|^2$. This will decrease the dependence of our LCSR on the value of the quark condensate and thus improve stability and reduce errors. We are left with the choice for the Borel window.

The Borel window is determined by two competing requirements. On the one hand, $t$ must be large enough to ensure that severing the twist expansion after twist 4 is valid, as the contributions of twist 5 and 6 are suppressed by an additional factor $\frac{1}{t}$ compared to the twist 3 and 4 contributions. On the other hand, a small Borel parameter is necessary to assure an adequate exponential suppression of the continuum and guarantees the validity of the quark-hadron duality. This suggests the interval 
\begin{equation}
\label{Borelwindow}
1{\rm{GeV}}^2 \; \leq t \leq 2 \; {\rm{GeV}}^2  \mbox{ .}
\end{equation}

\begin{figure}[t]
\begin{center}
\vspace{5mm}
\epsfig{figure=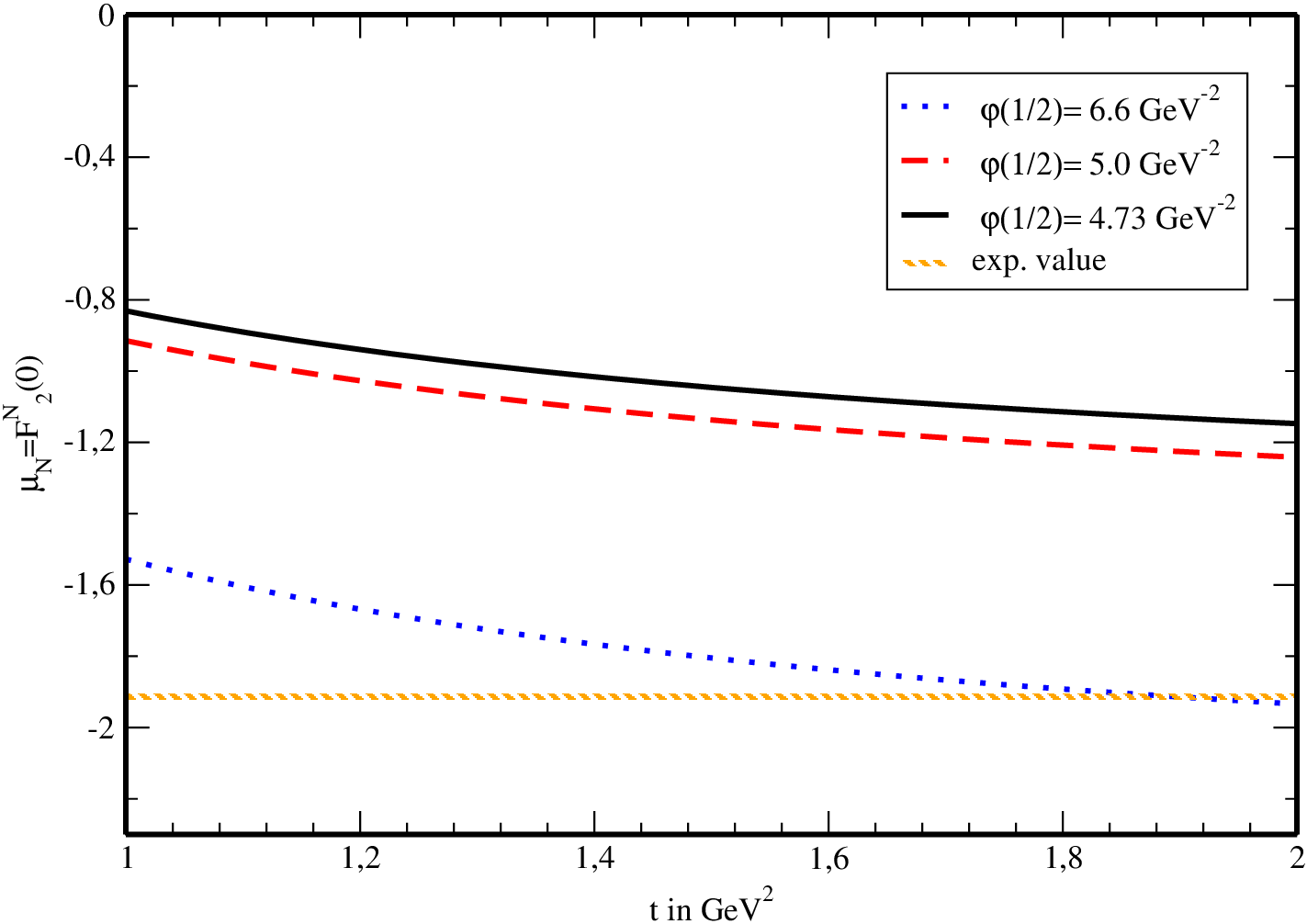 ,width=0.8\textwidth} 
\end{center}
\caption{ Magnetic moment of the neutron from sum rule \eqref{finalSR} for different values of $\varphi(1/2)$. The hatched line represents the experimental value \cite{PDG}.   \label{NeutronGraph}}
\end{figure}

In Fig.\ref{ProtonGraph} we plotted the sum rule \eqref{finalSR} for different values of $\varphi(1/2)$. The comparison with the experimental value \cite{PDG} $\mu_P\approx2.793$ favours a value $\varphi(1/2)=5.25 \pm 0.15 \;{\rm GeV^{-2}}$. This corresponds to $\chi(\mu=1\;{{\rm GeV}})=3.5\pm 0.1 \;{\rm GeV}^{-2}$, assuming one uses the asymptotic expression for the DAs. This agrees rather well with the result from the vector dominance model and the latest QCD sum rule result. Note that a larger value of $\chi$ can still be realized, if the full DA has a rather flat shape, this is the case e.g. in the instanton-model of the QCD vacuum \cite{Petrov:1998kg, Praszalowicz:2001wy}. 

The magnetic moment of the neutron can, as already stated, be obtained from Eq.\eqref{finalSR} by exchanging $e_u \leftrightarrow e_d$. We plotted the results in Fig. \ref{NeutronGraph}. As in \cite{Braun:1988qv}, the sum rule prediction for $\mu_N$ is somewhat below the experimental value \cite{PDG} $\mu_N\approx-1.913$.

The experimental result for the neutron magnetic moment $\mu_N$ can be reproduced by using a $40\%$ larger value of $\phi(1/2)$ that can be achieved if the magnetic susceptibility $\chi$ is increased by the corresponding amount, or if the photon distribution amplitude is more peaked in the middle point as might be suggested by the model for the distribution amplitude for transversely polarized mesons in \cite{Bakulev:2000er}. The stability of the neutron sum
rule becomes, however, somewhat worse in this case. Also also the agreement of the sum rule prediction for the proton magnetic moment is spoiled. Also $\mu_N$ turns out to be more sensitive than $\mu_p$ to those higher twist corrections that are only known to circa $50\%$ accuracy so that the overall error is larger. Worse agreement for $\mu_N$ compared to $\mu_P$ is therefore no surprise.

With the standard choice $\varphi(u) \approx \varphi^{asy}(u)$ and the latest value  $\chi(\mu=1\;\!{\rm{GeV}}^{2})=3.15\pm 0.3 \;\!{\rm{GeV}}^{-2}$ \cite{Ball:2002ps} \footnote{$\chi$ only appears in the combination $\qcond \chi$, which has a very weak scale dependence, see e.g. \cite{Ball:2002ps}. Thus the evolution effect is negligible.}  we obtain the following results for the magnetic moments
\begin{align}
\mu_P=& \phantom{(}  1+ (0.96 \pm 0.1)_{tw-2} + (0.72 \pm 0.18)_{tw-4} \nn \\
=&\phantom{(}2.68 \pm 0.28\\
\mu_N=&          (-1.93 \pm 0.2)_{tw-2} + (0.88 \pm 0.27)_{tw-4} \nn \\
=&\phantom{(}1.05 \pm 0.47 \mbox{.}
\end{align}
Here we also included the different contributions of twist 2 and twist 4 to the magnetic moments and how the overall error is distributed.

In the following we will stick to the above standard choice for $\varphi(u)$.

\section{The Nucleon-Delta-transition}

We will now expand the technique presented in the previous Section to the case of the nucleon-$\Delta$-transition. We will not go into the details of the calculation, except for the construction of a suitable Lorentz basis. This enables us to remove unwanted contributions of transitions including final states with isospin $3/2$ and spin $1/2$. 

It is not necessary to treat the the $n\gamma \to \Delta^{0}$ transition separately from the $p\gamma \to \Delta^{+}$ transition, as the final formulae will only differ by the exchange $e_u \leftrightarrow e_d$. Hence, we will only consider the proton transition.

\subsection{Definitions}
In order to study the $p \gamma \to \Delta^{+}$ transition using LCSRs, it is convenient to consider the correlation function corresponding to the diagram in Fig.\ref{PtoPgamma}:

\begin{figure}[t]
\centering{\epsfig{figure=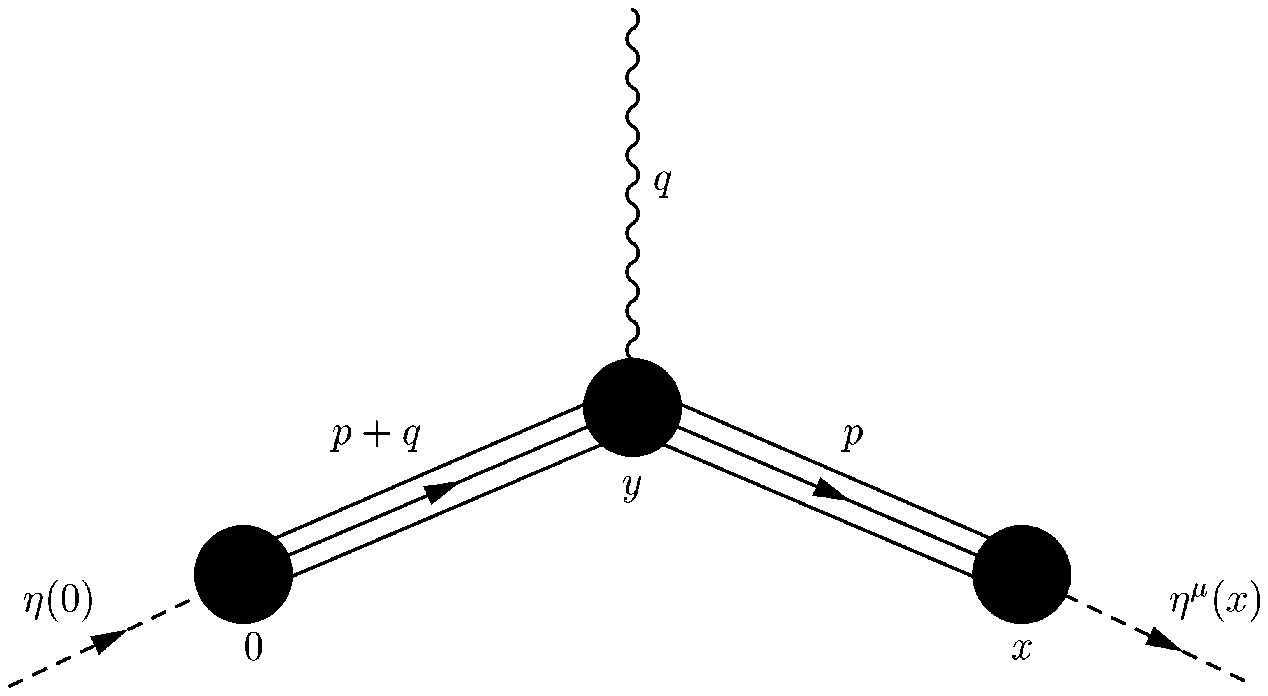,width=0.6\textwidth}}
\caption{Graph corresponding to Eq.\eqref{correlatorPreBackground} in coordinate space.  \label{PtoPgamma}}
\end{figure}

\begin{equation}
\label{correlatorPreBackground}
\Pi^{\mu \nu}(p,q) =i^2\int \!\! d^4x \int \!\! d^4y\; e^{ipx+iqy} \bra{0}\mathcal{T} \left \lbrace  \eta^{\mu}(x) j^{\nu}(y) \bar{\eta}(0) \right \rbrace \ket{0} \mbox{ .}
\end{equation}
For the current $\eta^{\mu}$ that creates states with the quantum numbers of the $\Delta^{+}$, we will follow a suggestion by Ioffe \cite{Ioffe:1981kw} and use 
\begin{align}
\label{DeltaCurrent}
\eta^{\mu}(x)=& \left[\left(u^a(x)\mathcal{C}\gamma^{\mu} u^{b}(x) \right) d^{c}(x)+2\left(u^a(x)\mathcal{C}\gamma^{\mu} d^{b}(x) \right) u^{c}(x) \right]\varepsilon^{abc} \mbox{ .}
\end{align}

Note that the current $\eta$, Eq.\eqref{IoffeCurrent}, has isospin $1/2$, whereas $\eta^{\mu}$ has isospin $3/2$, therefore only the isovector part of the electromagnetic current 
 \begin{equation}
\label{EmCurrentIsoVector}
j_{\mu}^{(I=1)}=\frac{1}{2}\left(e_u-e_d \right) \left(\ubar \gamma_{\mu} u - \dbar \gamma_{\mu} d \right)
\end{equation}
can induce the $p \gamma \to \Delta^{+}$ transition \cite{Belyaev:1995ya}. 

We will also use a correlation function in an electromagnetic background field $F_{\mu \nu}$
\begin{equation}
\label{correlator}
\Pi^{\mu \nu}_{\Delta}(p,q) e^{\left( \lambda \right)}_ {\nu}=i\int \!\! d^4x\; e^{ipx} \bra{0}\mathcal{T} \left \lbrace  \eta^{\mu}(x) \bar{\eta}(0) \right \rbrace \ket{0}_F \mbox{ .}
\end{equation}

The quantities, which can be measured experimentally, are the magnetic dipole transition form factor $G_{M}(0)$ and the electric quadrupole transition form factor $G_{E}(0)$. They are given by \cite{Jones:1972ky}
\begin{align}
\label{GM}
G_M(0)=&\frac{m_P}{3\left(m_P+m_{\Delta}\right)} \bigg[  \left( 3m_{\Delta} +m_P \right) \left( m_{\Delta}+m_P \right) \frac{G_1(0)}{m_{\Delta}}\nn \\
&\phantom{\frac{m_P}{3\left(m_P+m_{\Delta}\right)} \bigg[} +\left(m_{\Delta}^2-m_P^2 \right) G_2(0) \bigg] \qquad \\
\label{GE}
G_E(0)=& \frac{m_P}{3\left(m_P+m_{\Delta}\right)} \bigg[ \left(m_{\Delta}^2-m_P^2 \right) \left(\frac{G_1(0)}{m_{\Delta}} + G_2(0)\right) \bigg]
\end{align}
 and 
\begin{align}
\label{REM}
R_{EM}=&-\frac{G_{E}(0)}{G_{M}(0)}
\end{align}
where $G_{i}(0)$, $i=1,2$ are form factors defined in Eq.\eqref{DeltaNVertex}.

\subsection{Contribution of $p \gamma \to \Delta^{+}$ to the correlation function}

The contribution of the $p \gamma \to \Delta^{+}$ transition to the correlator 
\eqref{correlatorPreBackground} is given by 
\begin{align}
\label{NDcontribution}
T^{\mu \nu}_{\Delta} e^{\left( \lambda \right)}_ {\nu}=&\frac{1}{m_{\Delta}^2-p^2} \frac{1}{m_{P}^2-(p+q)^2} \sum_{s,s'}\bra{0} \eta_{\mu}(0) \ket{\Delta(p,s)} \times \nn \\& \bra{\Delta(p,s)} j^{\nu}(0) \ket{P(p+q,s')} \bra{P(p+q,s')} \overline{\eta}(0)\ket{0} e^{(\lambda)}_{\nu} \mbox{ .}
\end{align}
The matrix element
\begin{align}
\label{DeltaCoupling}
\bra{0} \eta_{\mu}(0)\ket{\Delta(p,s)}=\frac{\lambda_{\Delta}}{(2\pi)^2} \Delta^{(s)}_{\mu}(p)
\end{align}
is parametrised via the coupling constant $\lambda_{\Delta}$.
Here $\Delta^{(s)}_{\mu}(p)$ is a Rarita-Schwinger spinor for the $\Delta$.
Inserting the general expression for the transition matrix element
\begin{equation}
\label{DeltaNTranstion}
\bra{\Delta^{\mu}(p,s)} j_{\nu}(0)\ket{P(p+q,s')}=\overline{\Delta}^{\mu}(p,s) \Gamma_{\mu \nu} \gamma_5 P(p+q,s') \mbox{ ,}
\end{equation}
where the vertex $\Gamma_{\mu \nu}$ defines the three form factors $G_{1}(Q^2)$, $G_{2}(Q^2)$ and $G_{3}(Q^2)$,
\begin{align}
\label{DeltaNVertex}
\Gamma_{\mu \nu}=& \phantom{+} G_1\left(Q^2 \right)\left(g_{\mu \nu} \slashed{q} - q_{\mu} \gamma_{\nu} \right) +G_2\left(Q^2 \right)\left( g_{\mu \nu} q\cdot\left(p+\frac{q}{2} \right)-q_{\mu} \left( p+\frac{q}{2}\right)_{\nu}  \right) \nonumber \\ & +G_3\left(Q^2 \right) \left(q_{\mu} q_{\nu}-q^2 g_{\mu \nu} \right) \mbox{ ,}
\end{align}
in Eq.\eqref{NDcontribution} and using the spin summation formula
\begin{align}
\label{spinsum2}
\sum_{s} \Delta^{(s)}_{\mu}(p) \overline{\Delta}^{(s)}_{\mu}(p) = &-(\slashed{p}-m_{\Delta}) \left[ g_{\mu \nu} - \frac{1}{3}\gamma_{\mu} \gamma_{\nu}-\frac{ 2 p_{\mu} p_{\nu} }{ 3m_{\Delta}^2 } + \frac{ p_{\mu}\gamma_{\nu}-p_{\nu}\gamma_{\mu} }{ 3 m_{\Delta} } \right]
\end{align}
the contribution $T^{\mu \nu}_{\Delta} e_{\nu}^{(\lambda)}$ can be written as
\begin{align}
\label{GoodContribution}
T^{\mu \nu}_{\Delta} e^{\left( \lambda \right)}_ {\nu}=&-\frac{\lambda_{\Delta} \lambda_{P}}{(m_{\Delta}^2-p^2)(m_{P}^2-(p+q)^2)}(\slashed{p}-m_{\Delta}) \times\nn\\ \ &\left[ g^{\mu}_{\alpha} - \frac{1}{3}\gamma^{\mu} \gamma_{\alpha}-\frac{ 2 p^{\mu} p_{\alpha} }{ 3m_{\Delta}^2 } + \frac{ p^{\mu} \gamma_{\alpha}-p_{\alpha} \gamma^{\mu} }{ 3 m_{\Delta} } \right]\Gamma^{\alpha \nu}\gamma_5 (\slashed{p}+\slashed{q}-m_{P}) \mbox{ .}
\end{align}
However, it is known that the correlation function is plagued 
by transitions $p \gamma \to N^*$\cite{ Belyaev:1995ya, Peters}, where $\ket{N^*}$ is a isospin-$\frac{3}{2}$ 
spin-${\frac{1}{2}}$ state with negative parity. It is advantageous to use the Lorentz structures that do not receive contributions due to transitions to spin-${\frac{1}{2}}^{-}$ states. To find these we consider the overlap of $\ket{N^*}$ with $\eta^{\mu}$, which is defined \cite{Peters} as
\begin{equation}
\bra{0} \eta_{\mu}(0)\ket{N^*(p,s)}=\frac{\lambda_{N^*}}{(2\pi)^2}(m_{*}\gamma^{\mu}-4p^{\mu}) N^{*{(s)}}(p) \mbox{ ,}
\end{equation}
where $\lambda_{N^*}$ is the coupling and $m_{*}$ the mass of the spin-$1/2$ state.
The spinor $N^{*{(s)}}(p)$ satisfies the Dirac equation $(\slashed{p}-m_{*})N^{*{(s)}}(p)=0$ and equation \eqref{spinsum1}. 
Using the general decomposition of the electromagnetic transition matrix element
\begin{align}
\label{spin1/2transition}
\bra{N^*(p,s)} j^{\nu}(0) \ket{N(p+q,s')}=& \overline{N^*}^{(s)}(p)\left[\left( \gamma^{\nu}q^2 - \slashed{q} q^{\nu} \right)F^{NN^*}_{1}\!(Q^2) \right. \nn \\
&\left.-i \sigma^{\nu \alpha} q_{\alpha} F^{NN^*}_{2}\!(Q^2) \right]\gamma_5 N^{(s')}(p+q) \mbox{ ,}
\end{align}
we can write the unwanted contribution to $\Pi^{\mu\nu}$ as
\begin{align}
\label{BadContribution}
T^{\mu\nu}_{(1/2)}=&\frac{\lambda_{N^*}\lambda_{N}}{(m^2_{N^*}-p^2)(m_N^2-(p+q)^2)} (m_{*}\gamma^{\mu}-4p^{\mu})(\slashed{p}-m_{N^*}) \times \nn \\&
\left[\left( \gamma^{\nu}q^2 - \slashed{q} q^{\nu} \right)F^{NN^*}_{1}(Q^2) -i \sigma^{\nu \alpha} q_{\alpha} F^{NN^*}_{2}(Q^2) \right]\gamma_5 (\slashed{p}+\slashed{q}-m_{N}) \mbox{ .}
\end{align}
In Refs.\cite{ Belyaev:1995ya, Peters} it has been shown that it is possible to disentangle the contribution to $\Pi^{\mu\nu}$ due to Eq.\eqref{BadContribution} and Eq.\eqref{GoodContribution} by a specific choice for the Lorentz basis. As our kinematics are different from those in Ref.\cite{Peters}, we cannot use the same basis, however.

\subsection{Lorentz basis}

The correlator Eq.\eqref{correlatorPreBackground} satisfies two independent constraints, which have to be taken into account when constructing a suitable Lorentz basis:
\begin{itemize}
\item the transversality condition $q_{\nu} \Pi^{\mu \nu}=0$
\item the Rarita-Schwinger condition $\gamma_{\mu} \Pi^{\mu \nu}=0$ .
\end{itemize}
As the transversality condition is automatically fulfilled by Lorentz structures proportional to $F^{\mu \nu}$ and its derivatives, it is convenient to construct a basis from these structures. This yields $20$ different Lorentz structures and the QED Bianchi Identity eliminates four thereof. The Rarita-Schwinger condition provides four additional constraints reducing the number of independent Lorentz structures to a mere 12:
\begin{align}
 \mathcal{R}_{1}=&\;qp\;\gamma^{\mu}\slashed{e}\slashed{p} \gamma_5-ep\;\gamma^{\mu}\slashed{q}\slashed{p} \gamma_5+4 \left(qp\;p^{\mu}\slashed{e} \gamma_5-ep\;p^{\mu}\slashed{q} \gamma_5 \right)
 \nn \\
\mathcal{R}_{2}=&\;p^2 \left(qp\;\gamma^{\mu}\slashed{e} \gamma_5-ep\;\gamma^{\mu} \slashed{q}\gamma_5 \right)+4 \left(qp\;p^{\mu}\slashed{e}\slashed{p} \gamma_5-ep\;p^{\mu} \slashed{q}\slashed{p}\gamma_5 \right)
\nn \\
\mathcal{R}_{3}= &\;qp\;\gamma^{\mu}\slashed{e} \gamma_5-ep\;\gamma^{\mu} \slashed{q}\gamma_5+2 \left(p^{\mu}\slashed{e}\slashed{q}\gamma_5 \right) -\frac{1}{2} \left( \gamma^{\mu}\slashed{e}\slashed{q}\slashed{p}\gamma_5\right)
\nn \\
\mathcal{R}_{4}=&\;4 \left(p^{\mu}\slashed{e}\slashed{q}\slashed{p}\gamma_5 \right)-p^2 \left( \gamma^{\mu}\slashed{e}\slashed{q} \gamma_5\right) +2 \left( qp\;\gamma^{\mu}\slashed{e}\slashed{p} \gamma_5-ep\;\gamma^{\mu} \slashed{q}\slashed{p}\gamma_5\right)
\nn \\
\mathcal{R}_{5}=& \;qp \left( \gamma^{\mu}\slashed{e}\slashed{q} \gamma_5 \right) -4\left(qp\;e^{\mu}\slashed{q} \gamma_5-ep\;q^{\mu} \slashed{q}\gamma_5 \right)
\nn \\
\mathcal{R}_{6}=&\;\gamma^{\mu}\slashed{e}\slashed{q} \gamma_5-2 \left(e^{\mu} \slashed{q}\gamma_5-q^{\mu} \slashed{e}\gamma_5 \right)
\nn \\
\mathcal{R}_{7}= &\;qp\;\gamma^{\mu}\slashed{e} \gamma_5-ep\;\gamma^{\mu} \slashed{q}\gamma_5-4 \left(  qp\;e^{\mu} \gamma_5-ep\;q^{\mu} \gamma_5\right)
\nn \\
\mathcal{R}_{8}=&\;4\left(qp\;e^{\mu}\slashed{q}\slashed{p} \gamma_5-ep\;q^{\mu}\slashed{q}\slashed{p}\gamma_5 \right) - qp \left(\gamma^{\mu}\slashed{e}\slashed{q}\slashed{p}\gamma_5 \right) 
\nn \\
\mathcal{R}_{9}=&\;2 \left(q^{\mu}\slashed{e}\slashed{p} \gamma_5-e^{\mu} \slashed{q}\slashed{p}\gamma_5 \right) +\gamma^{\mu}\slashed{e}\slashed{q}\slashed{p}\gamma_5
\nn \\
\mathcal{R}_{10}=&\;4 \left( qp\;e^{\mu}\slashed{p} \gamma_5-ep\;q^{\mu} \slashed{p}\gamma_5 \right)- qp\;\gamma^{\mu}\slashed{e}\slashed{q} \gamma_5+ep\;\gamma^{\mu}\slashed{q}\slashed{p} \gamma_5
\nn \\
\mathcal{R}_{11}=&\;q^{\mu}\slashed{e}\slashed{q}\slashed{p}\gamma_5
\nn \\
\label{BASIS}
\mathcal{R}_{12}=&\;q^{\mu}\slashed{e}\slashed{q}\gamma_5 \mbox{.}
\end{align}
Here we inserted the explicit expression for $f_{\mu \nu}$, see Eq.\eqref{Fmunu}, which is advantageous for the further calculation.

\subsubsection{The $p \gamma \to \Delta^{+}$ contribution}

The contribution of the nucleon-$\Delta$-transition to the correlation function, Eq.\eqref{correlatorPreBackground}, can be obtained readily from Eq.\eqref{NDcontribution}. Using the basis \eqref{BASIS}, the result has the form:
\begin{align}
\label{GoodTerm}
T^{\mu \nu}_{\Delta}(p,q) e^{\left( \lambda \right)}_{\nu} =& \;- \lambda_{\Delta} \lambda_P  \frac{1}{\left(m_{\Delta}^2-p^2 \right) \left(m_{P}^2-(p+q)^2 \right)} \
\nonumber \\
&\times\left[ \left( \frac{(p+q)^2 G_{1}(0) }{96 m_{\Delta}^2 \pi^4}-\frac{m_{P} G_{2}(0) }{192 \pi^4}+ \frac{(p+q)^2 G_{2}(0) }{192 m_{\Delta} \pi^4} \right)\cdot  \mathcal{R}_{1} \right. 
\nonumber \\
&+ \left(  -\frac{m_{P} G_{1}(0)}{96 m_{\Delta}^2 \pi^4}+\frac{ G_{2}(0)}{192 \pi^4}-\frac{m_{P} G_{2}(0)}{192 m_{\Delta} \pi^4}\right)\cdot  \mathcal{R}_{2}
\nonumber \\
&+\left( -\frac{m_{P}G_{1}(0)}{48 \pi^4}-\frac{p^2 G_{1}(0)}{48 m_{\Delta} \pi^4}+\frac{2qp G_{2}(0)}{192 \pi^4}-\frac{2qp G_{1}(0)}{48 m_{\Delta} \pi^4} \right)\cdot  \mathcal{R}_{3}
\nonumber \\
&+\left( \frac{G_{1}(0) }{96 \pi^4}+\frac{m_{P}G_{1}(0) }{96 m_{\Delta} \pi^4} - \frac{2qpG_{1}(0) }{192 m_{\Delta}^2\pi^4}-\frac{2qpG_{2}(0) }{384 m_{\Delta} \pi^4} \right) \cdot  \mathcal{R}_{4}
\nonumber \\
&+\left( \frac{G_{1}(0)}{32 \pi^4}+\frac{m_{\Delta} G_{2}(0)}{64 \pi^4} \right) \cdot  \mathcal{R}_{5}
\nonumber \\
&+\left( -\frac{m_{P}m_{\Delta}G_{1}(0)}{32 \pi^4}-\frac{p^2G_{1}(0)}{32 \pi^4} -\frac{2qpG_{1}(0)}{32 \pi^4}  \right)\cdot  \mathcal{R}_{6}
\nonumber \\
&+\left( -\frac{m_{P}G_{1}(0)}{32 \pi^4}-\frac{m_{P}m_{\Delta}G_{2}(0)}{48 \pi^4}+\frac{p^2 G_{2}(0)}{48 \pi^4}+\frac{2qpG_{2}(0)}{48 \pi^4} \right)\cdot  \mathcal{R}_{7}
\nonumber \\
&+ \left( \frac{G_{2}(0)}{64 \pi^4} \right) \cdot  \mathcal{R}_{8}
\nonumber \\
&+ \left( \frac{m_{P} G_{1}(0)}{32 \pi^4}+\frac{m_{\Delta} G_{1}(0)}{32 \pi^4}\right)  \cdot  \mathcal{R}_{9}
\nonumber \\
&+ \left( -\frac{G_{1}(0)}{32 \pi^4} +\frac{m_{P}G_{2}(0)}{48 \pi^4}-\frac{m_{\Delta}G_{2}(0)}{48 \pi^4} \right)  \cdot  \mathcal{R}_{10}
\nonumber \\
&+ \left(\frac{G_{1}(0) }{16 \pi^4} \right)\cdot \mathcal{R}_{11} 
+\left.  \left( \frac{m_{\Delta} G_{1}(0)}{16 \pi^4} \right)\cdot \mathcal{R}_{12}\;\hspace{0.3mm}\right]
\end{align}

\subsubsection{The $N \gamma \to N^{*}$ contribution}

Analogously one finds for the contribution of the $J^P={\frac{1}{2}}^{-}$ states to Eq.\eqref{correlatorPreBackground}:
\begin{align}
\label{EvilTerm}
T^{\mu \nu}_{1/2}(p,q)e^{(\lambda)}_{\nu}=&- \lambda_{N^{*}} \lambda_N  \frac{1}{\left(m_{*}^2-p^2 \right) \left(m_P^2-(p+q)^2 \right)}
 \nn  \\
&\times\bigg[4 m_P F_2^{*}(0)\cdot \mathcal{R}_{1}-4 F_2^{*}(0)\cdot \mathcal{R}_{2}
\nn \\
 &\phantom{\times\bigg[}+\left( 4 p^2 F_2^{*}(0) - 4 m_P m_{*}F_2^{*}(0) \right)\cdot \mathcal{R}_{3} 
\nn \\
&\phantom{\times \bigg[}+\left(2m_{*}F_2^{*}(0) - 2m_P F_2^{*}(0) \right) \cdot \mathcal{R}_{4} \bigg] \mbox{ .}
\end{align}
We see that only the Lorentz structures $\mathcal{R}_1$, $\mathcal{R}_2$, $\mathcal{R}_3$ and $\mathcal{R}_4$ receive spin-1/2-contributions. In addition to that, the coefficient of the linear combination
\begin{equation}
\mathcal{R}_{1}+m_{P} \mathcal{R}_{2}
\end{equation}
vanishes for the $N \gamma \to N^{*}$ transition. Thus, we can choose among 9 structures which are suitable for the two sum rules for $G_1(0)$ and $G_2(0)$.

\subsection{Sum rules and numerical results}

The light-cone expansion of the correlation function is completely analogous to Section \ref{Proton}, even the Feynman diagrams are identical, see Fig.\ref{FD}. 
The calculation can be simplified further if the isospin relation requiring an overall factor $(e_u-e_d)$ is taken into account. It is then sufficient to calculate only terms proportional to $e_d$, which are simpler than those proportional to $e_u$.
The lengthy results for the various diagrams can be found in Appendix \ref{B}.

Before writing down sum rules for $G_1(0)$ and $G_2(0)$, it is advisable to identify those Lorentz structures promising the most reliable results. We will use two criteria:
\begin{enumerate}
\item the structure is free of spin-1/2 contributions 
\item the structure has the highest possible power of the momentum $p$.
\end{enumerate}
These demands are fulfilled  by $$\mathcal{R}_{5}= \;qp \left( \gamma^{\mu}\slashed{e}\slashed{q} \gamma_5 \right) -4\left(qp\;e^{\mu}\slashed{q} \gamma_5-ep\;q^{\mu} \slashed{q}\gamma_5 \right) \mbox{ ,}$$  $$\mathcal{R}_{8}=4\left(qp\;e^{\mu}\slashed{q}\slashed{p} \gamma_5-ep\;q^{\mu}\slashed{q}\slashed{p}\gamma_5 \right) - qp \left(\gamma^{\mu}\slashed{e}\slashed{q}\slashed{p}\gamma_5 \right)\mbox{ ,}$$ and
$$\mathcal{R}_{11}=q^{\mu}\slashed{e}\slashed{q}\slashed{p}\gamma_5 \mbox{ .}$$
Upon collecting the corresponding terms from Eqs.\eqref{GoodTerm},\eqref{Twist1},\eqref{Twist2},\eqref{PhotonResult} and \eqref{AllGluon}, one can easily assemble the sum rules corresponding to the three structures. 

The necessary Borel transformation can again be  performed with Eqs.\eqref{PBorel1}, \eqref{PBorel2}. There is, however, a subtlety: the choice of the ratio of the two Borel parameters $M_1^2$ and $M_2^2$.
If we were able to calculate the correlation functions exactly, the dependence on the Borel parameters would vanish, as they are not physical quantities. Our calculation is, however, approximate, and the approximation is rather crude so that the difference in the mass scales in the $\Delta$ and the nucleon channels is not reproduced by the sum rules. This can be checked by the calculation of the masses in the two momentum channels using standard techniques. This mass difference between the nucleon and the $\Delta$ is large, of order $300 \; {\rm MeV}$, and it has to be taken into account. Neglecting this difference, from our point of view, is the main reason why the Ioffe-Smilga sum rule for the nucleon $\Delta$ magnetic transition did not produce acceptable results. In the approach that we are using there is a possibility to take into account the mass difference because the two momenta $p_1^2$ and $p_2^2$ alias Borel parameters $M_1^2$  and $M_2^2$ are taken as independent variables, and the sum rule can be ``repaired'' by taking the ratio of Borel parameters at a fixed value 
\begin{equation}
\label{Borelratio}
    \frac{M_1^2}{M_2^2} = \frac{m_{\Delta}^2}{m_N^2} \mbox{ .}
\end{equation}
Since this ratio determines the momentum fraction ratio, at which e.g. the photon wave function $\varphi(u)$ is evaluated, this choice shifts $u$ away from the centre $1/2$. We will follow this strategy, which was advocated in \cite{Balitsky:1989ry} for the calculation of the asymmetry in the $\Sigma \to p\gamma$ decay. A similar trick is often used in the calculation of SU(3) flavour symmetry breaking effects in the sum rule method.

The final sum rules for $G_1(0)$ and $G_2(0)$ are given by
\begin{align}
\label{G_1SR}
G_1(0)&=16 \pi^2\frac{(e_u-e_d)\qcond}{\lambda_{\Delta} \lambda_P} e^{M^2/t} \left \lbrace \frac{v}{6} \varphi(v) t^2  \left( 1-e^{-S_0/t} \left( 1+\frac{S_0}{t} \right) \right)+\mathcal{I}_1 \right \rbrace
\end{align}
and
\begin{align}
\label{G_2SR}
G_2(0)=&-64 \pi^4 \frac{e_u-e_d}{\lambda_{\Delta} \lambda_P }e^{M^2/t} \bigg \lbrace \frac{1}{64 \pi^4}v \overline{v} t^2 \left(1-e^{-S_0/t}\left(1+\frac{S_0}{t} \right)\right) +\mathcal{I}_2\bigg \rbrace
\end{align}
and 
\begin{align}
\label{G_1+G_2SR}
G_1(0)&+\frac{m_{\Delta}}{2}G_2(0)= 8 \pi^2 \frac{(e_u-e_d)\qcond}{\lambda_{\Delta} \lambda_P }e^{M^2/t}\bigg [ \frac{v \mathds{B}(v)}{4} t (1-e^{-S_0/t}) +\mathcal{I}_3 \bigg]
\end{align}
with
\begin{align}
M^2=&\frac{2m_{\Delta}^2 m_p^2}{m_p^2+m_{\Delta}^2}\\
S_0=&\frac{2s_{\Delta}^2 s_{P}^2}{s_{\Delta}^2 +s_{P}^2}\\
v=&\frac{M_1^2}{M_2^2 + M_1^2} \mbox{ .}
\end{align}
 $\mathcal{I}_1$ and $\mathcal{I}_3$ correspond to lengthy contributions of twist 4, whereas $\mathcal{I}_2$ is of twist 3. The full expressions for $\mathcal{I}_1$, $\mathcal{I}_2$ and $\mathcal{I}_3$ can be found in Appendix \ref{B2}.
The magnetic dipole form factor $G_{M}(0)$ and the electric quadrupole form factor $G_{E}(0)$ can be obtained via Eqs.\eqref{GM}, \eqref{GE}.

In the following analysis we use the asymptotic expression for the leading photon wave function $\varphi(u)$, which yielded a reasonable result for the proton magnetic moment in Section 2.
The continuum threshold for the $\Delta$, $s_{\Delta}\approx 3.0\;\!{\rm{GeV}}^2$, can be determined from the sum rule for the coupling $|\lambda_{\Delta}^2|$ \cite{Belyaev:1982sa}. By using the whole sum rule expressions for $\lambda_{\Delta}$ and $\lambda_{P}$, the stability of the sum rule can again be improved.
The same Borel window as in Eq.\eqref{Borelwindow} is used.

\begin{figure}
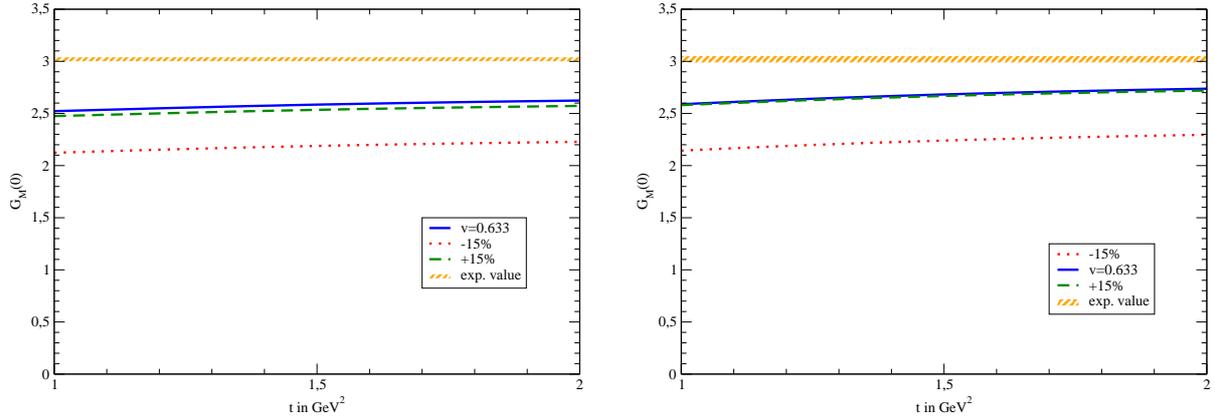

	\centering{	\epsfig{figure=GM1.eps,width=0.47\textwidth} \mbox{\;\;\;\;}
			\epsfig{figure=GM2.eps,width=0.47\textwidth} }
\caption{The magnetic form factor $G_{M}(0)$ of the $p\gamma \to \Delta^+$ transition. The solid black curve shows the result of the corresponding sum rule, the dashed and dotted curves arise from a variation of the ratio of Borel parameters by $15\%$. The experimental value \cite{Tiator:2000iy} including uncertainties is given by the hatched region. The left panel shows the result using $G_{2}(0)$ determined from sum rule Eq.\eqref{G_2SR}, the right panel using $G_{2}(0)$ determined from Eq.\eqref{G_1+G_2SR}. \label{GMPlot}}
\end{figure}

In Fig.\ref{GMPlot} our result for $G_{M}(0)$ is shown. 
In addition to our sum rule, we also plotted the results for a slightly changed ratio of the Borel parameters ($\pm 15 \%$). The result implies that the sum rules for $G_{M}(0)$ are rather stable with respect to a variation of $v$, if the Borel parameters are chosen as in \eqref{Borelratio}. The numerical values obtained if one determines $G_2(0)$ from Eq.\eqref{G_2SR} or Eq.\eqref{G_1+G_2SR}, respectively,
\begin{align}
G_{M}(0) &= (1.55 \pm 0.15)_{tw-2}+(-0.04 \pm 0.01)_{tw-3}+(1.08 \pm 0.2)_{tw-4} \nn  \\
&= 2.59 \pm 0.36 \\
G_{M}(0) &= (1.55 \pm 0.15)_{tw-2}+(1.15 \pm 0.12)_{tw-4}\nn  \\ &=2.70 \pm 0.27
\end{align}
are close together and agree rather well with experiment \cite{Tiator:2000iy}
\begin{equation}
G_M(0)= 3.02 \pm 0.03 \mbox{ .}
\end{equation}	
Note that the error in the leading twist contribution stems almost exclusively from the uncertainty of the value of the magnetic susceptibility $\chi$.

\begin{figure}[t]
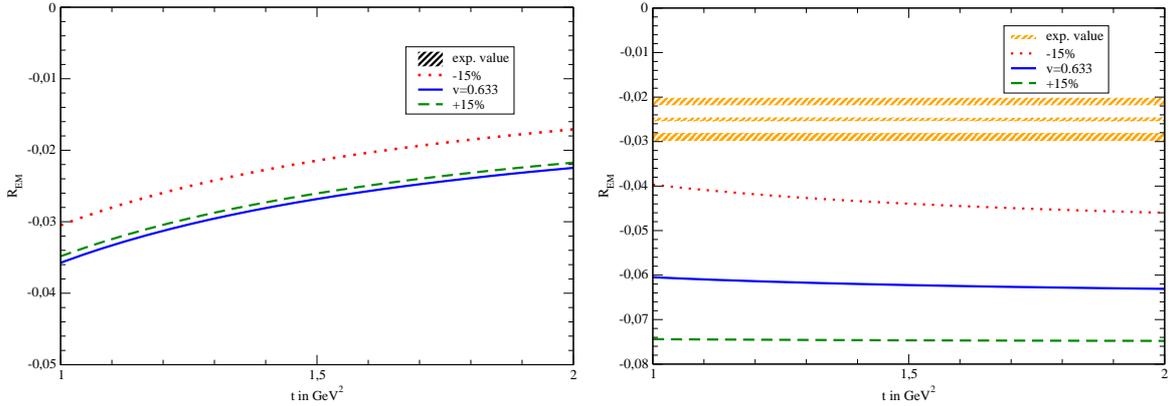

	\centering{\epsfig{figure=REM1.eps,width=0.47 \textwidth}\mbox{  }
		   \epsfig{figure=REM2.eps,width=0.47 \textwidth}}
	\caption{The ratio $R_{EM}$. The identification of the panels follows Fig.\ref{GMPlot}. The experimental value is from \cite{PDG}. The left panel shows the result using $G_{2}(0)$ determined from sum rule Eq.\eqref{G_2SR}, the right panel using $G_{2}(0)$ determined from Eq.\eqref{G_1+G_2SR}. \label{REMPlot}}
\end{figure}

Our estimates for the ratio $R_{EM}$ that can be obtained by using Eq.\eqref{REM} are shown in Fig.\ref{REMPlot}.
The results are 
\begin{align}
\label{littlemoretrust}
R_{EM}(0)&=(-7.6\%\pm 0.1\%)_{tw-2} + (1.15\% \pm 0.7\%)_{tw-3} +(0.05 \% \pm 0.02\% )_{tw-4} \nn \\
&\approx-(6.4 \pm 0.8)\%\\
\label{lesstrust}
R_{EM}(0)&=(-6.3\% \pm 0.35\%)_{tw-2} + (3.7\% \pm 0.25\%)_{tw-4}\nn \\
 &\approx -(2.8 \pm 0.6)\%\mbox{ ,}
\end{align}
where $G_{2}(0)$ is determined from the sum rule Eq.\eqref{G_2SR} and Eq.\eqref{G_1+G_2SR}, respectively. 
These values have to be compared to 
\begin{equation}
 R_{EM}(0)=(-2.5\pm 0.5)\% \mbox{ ,}
\end{equation}
given by the Particle Date Group \cite{PDG}.
Although the result \eqref{lesstrust} is closer to experimental data, it is less reliable, as Eq.\eqref{G_1+G_2SR} has no leading-twist contribution. This is also the reason, why the relative error in \eqref{lesstrust} is much larger than in \eqref{littlemoretrust}.
The agreement of both results is still very reasonable, taking into account the smallness of $R_{EM}$ that is largely due to cancellations.
It is therefore not unexpected that two different sum rules for the same quantity agree only within a factor of two and do not contradict the validity of our approach.

In \cite{Aliev:2004ju} the form factors of the nucleon-$\Delta$ transition have been calculated as part of a very general examination of radiative decays of decuplet baryons into octet baryons. This calculation has been carried out in LCSR and is, in principle, very similar to our calculation. Apart from some technical differences, such as a different choice for the Lorentz basis, there is one point that has to be addressed.
We explicitly included the electromagnetic background field in the quark propagator \eqref{propagator}. As discussed in \cite{Ball:2002ps}, working in this background field simplifies the treatment of the notorious contact terms.
In particular, this procedure allows to include in a natural way the contributions from photon distribution amplitudes, that are known to vanish exactly, but have a non-zero conformal expansion to next-to-leading order. These have to be taken into account as most photon DAs are only known to next-to-leading order in conformal spin, which requires all photon DAs to be of this accuracy. In \cite{Aliev:2004ju} such terms were neglected. The numerical impact of these contributions apparently is small and the final results for $G_M(0)=2.5 \pm 1.3$ and $R_{EM}(0)=-6.8\%$ from Ref.\cite{Aliev:2004ju} are  close to ours. This consistency lends support to the general technique of LCSRs using photon DAs.

\section{Conclusions}

We have calculated the nucleon magnetic moments and the electromagnetic transition form factors of $\gamma p \to \Delta^+$ for real photons using the light-cone sum rule approach. 
Our result for the magnetic moment of the proton $\mu_P$ is in good agreement with experiment and lends support to the current models of photon DAs and the estimates of the magnetic susceptibility of the quark condensate. The sum rule for the neutron magnetic moment is more sensitive to higher twist terms and therefore less accurate.

The calculation of the $\gamma p \to \Delta^+$ transition form factors is, in principle, analogous to the calculation of the nucleon magnetic moments. The main difference is the asymmetric choice of the Borel parameters, that allowed us to take the mass difference between proton and $\Delta^+$ into account. Refraining from doing so would lead to distinctly worse results. Within uncertainties, the magnetic dipole form factor $G_M(0)$ of the $p \gamma \to \Delta^+$ transition agrees well with current data. This result is rather surprising as a different approach \cite{Peters} also using light-cone sum rules, that is valid for $Q^2 > 1 \; {\rm GeV}^2 $ predicts a value for $G_M$ that is below data in the region $Q^2 <2  \; {\rm GeV}^2$. In order to close the gap to this calculation it is necessary to expand our approach from the real photon point to virtualities ranging from $0$ to $-1 {\rm GeV}^2$. This requires photon distribution amplitudes for virtual photons, see e.g. \cite{Yu:15}. 

Our results for the ratio $R_{EM}$ agree with experiment within a factor of $2$. The two different sum rules written down for $R_{EM}$ also differ from each other by a factor of $2$, while they agree very well for $G_M(0)$. This supports our presumption that a lower accuracy for $R_{EM}$ is due to considerable cancellations, so this quantity is intrinsically more difficult to calculate with precision. Both $G_{M}(0)$ and $R_{EM}(0)$ are in good agreement with the corresponding results from \cite{Aliev:2004ju}.

The best agreement for $\mu_P$ with experiment could be archived by the choice $$\varphi(1/2)= 5.25 \pm 0.15 {\rm GeV}^{-2}$$ and this value would also be favoured by the two sum rules for $G_M(0)$, provided the asymptotic shape for $\varphi(u)$ is used. The value of $\varphi(1/2)$ is an independent piece of information compared to the expansion in Gegenbauer polynomials, which can be obtained from the sum rules. However, this is not sufficient to provide insights into the shape of the distribution amplitude. 

In order to increase the accuracy of our calculation it is necessary to take into account $\alpha_s$ corrections. As the main source for the uncertainties are the numerical values of the twist-3 and twist-4 DAs that are known only up to at best $50\%$, it is an important task of its own to determine their values more precisely. This would improve the numerics in this paper and be valuable for future work.

\hspace{2cm}

\section*{Acknowledgements}
I am very grateful to V. Braun for many enlightening and clarifying discussions and to A. Lenz for proofreading the manuscript. I also would like to thank T. Aliev for pointing out the work of Ref.\cite{Aliev:2004ju}.
This work is supported by DFG -- 9209070.
\newpage

\begin{appendix}
\section{Photon distribution amplitudes}
\label{A}
%copy and paste
For completeness we collect the relevant photon distribution amplitudes for the $p \gamma \to \Delta$ transition according to \cite{Ball:2002ps}. Note that in \cite{Ball:2002ps} the photon momentum has the opposite sign and the parametrisation of the separation of antiquark and quark is different.

The path-ordered exponents (cp. Eq.\eqref{pexp})
\begin{equation*}
\left[ x,y \right]={\rm Pexp}\left \lbrace i \int \! dt \;(x-y)_{\mu} \left[ e_q A^{\mu}(tx-\overline{t}y) + g B^{\mu}(tx-\overline{t}y) \right] \right \rbrace
\end{equation*}
assure gauge invariance of the matrix elements. It is important, that the electromagnetic field is included herein, as additional terms to those given in \cite{Ball:2002ps} will occur otherwise.

\subsection{Twist-2 and Twist-4 DAs}
The leading-twist DA reads
 \begin{align}
\label{SigmaMatrixelement}
\bra{0} \qbar(0) \left[ 0,x \right]\sigma_{\alpha \beta} q(x) \ket{0}_F&=e_q \left\langle \qbar q \right\rangle \int _0 ^1  \! du \; \varphi (u) F_{\alpha \beta}(ux) \nn \\
     & \phantom{=} +\frac{e_q \left\langle \qbar q \right\rangle}{16} \int_0^1 \!du\; x^2 \mathds{A}(u) F_{\alpha \beta}(ux) \nn \\
     & \phantom{=} +\frac{e_q \left\langle\qbar q \right\rangle}{8}\int_0^1 \!du\;\mathds{B}(u)x^{\rho}\left(x_{\beta}F_{\alpha \rho}(ux) -x_{\alpha}F_{\beta \rho}(ux) \right)
\end{align}
with 
\begin{align}
\varphi(u)&=\varphi^{{\rm asy.}}(u)=6 \chi u(1-u) \\
\mathds{A}(u)&=40 u (1-u)\left(3\kappa -\kappa^{+}+1 \right)+8\left(\zeta^{+}_2-3\zeta_2 \right) \times 
\nn \\
 &\phantom{=} \left[ u(1-u)\left( 2+13u(1-u) \right)+2u^3\left(10-15u+6u^2 \right)\ln (u)\right.
\nn \\
&\phantom{=} \left.+ 2(1-u)^3\left(10-15(1-u)+6{1-u}^2 \right)\ln (1-u)\right]
 \\
\mathds{B}(u)&=40 \int_0^u \! d\alpha \; (u-\alpha) \left(1+3 \kappa^+ \right)\left[ -\frac{1}{2}+\frac{3}{2} (2\alpha -1)^2 \right] \mbox{ .}
\end{align}

\begin{align}
\label{photonDA1}
\bra{0}& \qbar(0) e_q \left[ 0,x \right] F_{\mu \nu} (ux) q(x) \ket{0}_F=e_q \qcond \int \Da \mathcal{S}_{\gamma}(\underline{\alpha})F_{\mu \nu}(\alpha_u x)\\
\label{photonDA2}
\bra{0}& \qbar(0) e_q \left[ 0,x \right] \sigma _{\alpha \beta} F_{\mu \nu} (ux) q(x) \ket{0}_F=
\nn\\
&-\frac{e_q \qcond}{qx} \left[ q_{\alpha} q_{\mu} e^{\lambda}_{\bot \nu} x_{\beta}-  q_{\beta} q_{\mu} e^{\lambda}_{\bot \nu}- q_{\alpha} q_{\nu} e^{\lambda}_{\bot \mu} x_{\beta} +q_{\beta} q_{\nu} e^{\lambda}_{\bot \mu} x_{\alpha} \right]\mathcal{T}^{\gamma}_4(u,qx)\\
\label{DistAmpl1}
\bra{0}&\qbar(0) \left[ 0,ux \right]g G_{\mu \nu }(ux)\left[ ux,x \right]q(x)\ket{0}_F=e_q \qcond \int\!\! \Da \mathcal{S}(\underline{\alpha}) F_{\mu \nu}(\alpha_u x)\\
\label{DistAmpl2}
\bra{0}&\qbar(0) \left[ 0,ux \right]i\gamma_5 g \widetilde{G}^{\mu \nu }(ux) \left[ ux,x \right] q(x)\ket{0}_F=e_q \qcond \int\!\! \Da \mathcal{\widetilde{S}}(\underline{\alpha}) F_{\mu\nu}(\alpha_u x)\\
\label{DistAmpl5}
\bra{0}&\qbar(0)\left[ 0,ux \right]\sigma_{\alpha \beta} g G_{\mu \nu}(ux) \left[ ux,x \right]q(x)\ket{0}_F=
\nn \\
 &= - e_q \qcond \left[ q_{\alpha} e ^{(\lambda)}_{\bot \mu}g_{\beta \nu}^{\bot}-q_{\beta}e ^{(\lambda)}_{\bot \mu}g_{\alpha \nu}^{\bot}-q_{\alpha}e ^{(\lambda)}_{\bot \nu}g_{\beta \mu}^{\bot}+q_{\beta}e ^{(\lambda)}_{\bot \nu}g_{\alpha \mu}^{\bot} \right] \mathcal{T}_1(u,qx)\quad\;\;\!
\nn\\
 & \phantom{=}- e_q \qcond \left[ q_{\mu} e ^{(\lambda)}_{\bot \alpha}g_{\beta \nu}^{\bot}-q_{\mu}e ^{(\lambda)}_{\bot \beta}g_{\alpha \nu}^{\bot}-q_{\nu}e ^{(\lambda)}_{\bot \alpha}g_{\beta \mu}^{\bot}+q_{\nu}e ^{(\lambda)}_{\bot \beta}g_{\alpha \mu}^{\bot} \right]\mathcal{T}_2(u,qx)\quad\;\;\!
\nn\\
 &\phantom{=}-\frac{e_q \qcond}{qx} \left[ q_{\alpha} q_{\mu} e^{\lambda}_{\bot \beta} x_{\nu}-  q_{\beta} q_{\mu} e^{\lambda}_{\bot \alpha}- q_{\alpha} q_{\nu} e^{\lambda}_{\bot \beta} x_{\mu} +q_{\beta} q_{\nu} e^{\lambda}_{\bot \alpha} x_{\mu} \right]\mathcal{T}_3(u,qx)
\nn\\
&\phantom{=}-\frac{e_q \qcond}{qx} \left[ q_{\alpha} q_{\mu} e^{\lambda}_{\bot \nu} x_{\beta}-  q_{\beta} q_{\mu} e^{\lambda}_{\bot \nu}- q_{\alpha} q_{\nu} e^{\lambda}_{\bot \mu} x_{\beta} +q_{\beta} q_{\nu} e^{\lambda}_{\bot \mu} x_{\alpha} \right]\mathcal{T}_4(u,qx)
\end{align}
Here we used 
\begin{eqnarray}
\int \Da &=& \int_0^1 \!d\alpha_q \; \int_0^1 \!d\alpha_{\qbar} \; \int_0^1 \!d\alpha_g \; \delta(1-\alpha_q-\alpha_{\qbar}-\alpha_g) \\
\alpha_u&=&\alpha_q+u\alpha_g\\
g_{\mu \nu}^{\perp}&=&g_{\mu \nu}-\frac{q_{\mu}x_{\nu}+q_{\nu}x_{\mu}}{qx}\\
e^{\perp\;\!(\lambda)}_{\mu}&=&g_{\mu \nu}^{\perp}\;\!e^{\nu\;\!(\lambda)}
\end{eqnarray}
and
\begin{align}
\mathcal{S}(\underline{\alpha})&=30\alpha_g^2 \left[ \left( \kappa +\kappa^+ \right)\left(1-\alpha_g \right)+ \left( \zeta_1+\zeta_1^+ \right)\left( 1-\alpha_g \right) \left( 1-2\alpha_g \right) \right.
\nn \\
&\phantom{=}\left.+\zeta_2 \left( 3\left( \alpha_{\qbar}-\alpha_q \right)^2-\alpha_g \left( 1-\alpha_g \right) \right) \right]
\\
\widetilde{\mathcal{S}} (\underline{\alpha})&=-30\alpha_g^2 \left[ \left( \kappa -\kappa^+ \right)\left(1-\alpha_g \right)+ \left( \zeta_1-\zeta_1^+ \right) \left( 1-\alpha_g \right) \left( 1-2\alpha_g \right)\right.
\nn \\
&\phantom{=}\left.+\zeta_2\left( 3\left( \alpha_{\qbar}-\alpha_q \right)^2-\alpha_g\left(1-\alpha_g \right) \right) \right]
\\
S_{\gamma} \left( \underline{\alpha} \right) &=60 \alpha_g^2\left( \alpha_{\qbar} + \alpha_q \right)\left( 4 - 7 \left( \alpha_q + \alpha_{\qbar} \right) \right) 
\\
\mathcal{T}_i(u,qx)&=\int \Da \; e^{i\alpha_u qx} T_i(\underline{\alpha})
\end{align}
with
\begin{align}
T_1(\underline{\alpha})&=-120 \left( 3 \zeta_2 +\zeta_2^+ \right)\left( \alpha_{\qbar}-\alpha_q \right) \alpha_{\qbar}\alpha_q \alpha_g
\\
T_2(\underline{\alpha})&=30 \alpha_g^2\left(\alpha_{\qbar}-\alpha_q \right) \left[\left(\kappa -\kappa^+ \right)+\left(\zeta_1-\zeta_1^+ \right)\left( 1-2\alpha_g \right) +\zeta_2 \left( 3-4 \alpha_g \right) \right]
\\
T_3(\underline{\alpha})&=-120 \left( 3 \zeta_2 -\zeta_2^+ \right)\left( \alpha_{\qbar}-\alpha_q \right) \alpha_{\qbar}\alpha_q \alpha_g
\\
T_4(\underline{\alpha})&=30 \alpha_g^2\left(\alpha_{\qbar}-\alpha_q \right) \left[\left(\kappa +\kappa^+ \right)+\left(\zeta_1+\zeta_1^+ \right)\left( 1-2\alpha_g \right) +\zeta_2 \left( 3-4 \alpha_g \right) \right]
\\
T^{\gamma}_4(\underline{\alpha})&=60 \alpha_g^2\left( \alpha_{\qbar}-\alpha_q \right)\left( 4 - 7 \left( \alpha_q + \alpha_{\qbar} \right) \right) \mbox{ .}
\end{align}
The abbreviation $\underline{\alpha}$ represents $(\alpha_q, \alpha_{\qbar}, \alpha_g)$.The values of the various constants can be found in table \ref{Numbers}.

It should be noted that the matrix element 
$$\bra{0} \qbar(0) e_q \left[ 0,x \right] \sigma _{\alpha \beta} F_{\mu \nu}(ux) q(x) \ket{0}_F$$ 
vanishes exactly if one sums up the whole conformal expansion. The expansion itself has, however, non-zero coefficients and thus in next-to-leading order in conformal spin the matrix element is different from zero. For the same reason the matrix element
$$ \bra{0} \qbar(0) e_q \left[ 0,x \right] F_{\mu \nu} (ux) q(x) \ket{0}_F $$
has herein mentioned form and not $e_q \qcond F_{\mu \nu} (ux) \mbox{ .}$

\subsection{Twist-3 DAs}

\begin{align}
\label{Twist3MatrixelementVektor}
\bra{0} \qbar(0)\left[ 0,x \right] \gamma_{\alpha} q(x) \ket{0}_F =-\frac{e_q}{2}f_{3 \gamma} \int_0 ^1\! du \; \overline{\psi}^{(V)}(u) x^{\rho}F_{\rho \alpha} 
\\
\label{Twist3MatrixelementAxial}
\bra{0} \qbar(0) \left[ 0,x \right]\gamma_{\alpha} \gamma_5 q(x) \ket{0}_F =-i\frac{e_q}{4}f_{3 \gamma} \int_0 ^1\! du \; \psi^{(A)}(u) x^{\rho}\widetilde{F}_{\rho \alpha} 
\\
\label{DistAmpl3}
\bra{0}\qbar(0) \left[ 0,ux \right]i g \gamma_{\alpha} G_{\mu \nu }(ux)\left[ ux,x \right]q(x)\ket{0}_F= \qquad\qquad\qquad
\nn \\
 =e_q f_{3\gamma}q_{\alpha} \left[q^{\nu} e^{(\lambda)}_{\bot \mu} - q^{\mu} e^{(\lambda)}_{\bot \nu}  \right] \int \Da \mathcal{V}(\underline{\alpha}) e^{i\alpha_uqx}\quad\\
\label{DistAmpl4}
\bra{0}\qbar(0) \left[ 0,ux \right] g \gamma_{\alpha}\gamma_5 \widetilde{G}_{\mu \nu }(ux)\left[ ux,x \right]q(x)\ket{0}_F=\qquad\qquad\qquad
\nn \\
=e_q f_{3\gamma}q_{\alpha} \left[q^{\nu} e^{(\lambda)}_{\bot \mu} - q^{\mu} e^{(\lambda)}_{\bot \nu}  \right] \int \Da \mathcal{A}(\underline{\alpha}) e^{i\alpha_uqx}\quad
\end{align}
Where
\begin{align}
\overline{\psi}^{(V)}(u)&=-20u(1-u)(2u-1)
\nn \\ 
&\phantom{=} +\frac{15}{16} \left(\omega_{\gamma}^A-3\omega_{\gamma}^{V} \right)u (1-u) (2u-1)\left(7(2u-1)^2-3 \right)\\
{\psi}^{(A)}(u)&=(1-(2u-1)^2) \left( 5 \left( 2u-1 \right)^2 -1\right)\frac{5}{2} \left( 1+\frac{9}{16}\omega_{\gamma}^{V}-\frac{3}{16} \omega_{\gamma}^{A}\right)\\
\mathcal{V}(\underline{\alpha})&=540 \omega_{\gamma}^V \left( \alpha_q - \alpha_{\qbar} \right) \alpha_q \alpha_{\qbar} \alpha_{g}^2\\
\mathcal{A}(\underline{\alpha})&=360\alpha_q \alpha_{\qbar} \alpha_g^2 \left[1+\omega_{\gamma}^A\frac{1}{2} \left( 7 \alpha_g-3 \right) \right] \mbox{ .}
\end{align}
\newpage
\subsection{Numerical values for the parameters at the renormalisation scale $\mu=1{\rm GeV}$}

\begin{table}[h]
\begin{center}
\begin{tabular}{c||c}
\hline
$\chi$ & $3.15\pm 0.3\;\!{\rm GeV}^{-2}$  \\\hline
$\kappa$ &  $0.2 $ \\\hline
$\kappa^+$ & $0  $ \\\hline
$\zeta_1$ & $0.4 $ \\\hline
$\zeta_1^+$ &$0  $ \\\hline
$\zeta_2$ & $0.3 $ \\\hline
$\zeta_2^+$ &$0  $ \\\hline
$f_{3\gamma}$ & $-(4\pm 2)\cdot 10^{-3}\;\!{\rm GeV}^2$      \\\hline
$\omega_{\gamma}^A$ &$-2.1\pm 1.0 $ \\\hline
$\omega_{\gamma}^V$ & $3.8\pm 1.8 $\\\hline
$\qcond$ &$-(240 \pm 10\;\!{\rm MeV} )^3$ \\\hline
\end{tabular}
\end{center}

\caption{Numerical values and uncertainties of the relevant parameters \cite{Ball:2002ps, Balitsky:1989ry} \label{Numbers}.} 
\end{table}

\section{Light-cone expansion of the $\gamma p \to \Delta^+$ correlation function}
\label{B}

Here we collect the results for the Feynman diagrams in Fig.\ref{FD} up to twist-4 accuracy.

\begin{itemize}
\item Diagram a: 
\begin{align}
\label{Twist1}
T_a(q,p)=&\;\frac{e_u-e_d}{\pi^4} \int \nolimits _0 ^1 \! du \; \left \lbrace \; \left( - \frac{1}{64} \ubar \right) \ln \left(\frac{\mu^2}{-\ubar p_1^2-u p_2^2} \right) \cdot \mathcal{R}_2 \quad\! \right. 
\nn \\
 &+\frac{1}{64} \left( up^2+3 u^2 p^2 +8 u^3 qp \right) \ln \left( \frac{\mu^2}{-\ubar p_1^2-u p_2^2} \right) \cdot \mathcal{R}_3 \quad 
\nn \\
& -\frac{1}{64} \left(3p^2+3 u p^2 +6 u qp -6 u^2 qp \right) \ln \left( \frac{\mu^2}{-\ubar p_1^2-u p_2^2} \right) \cdot \mathcal{R}_7 \quad 
\nn \\
& -\frac{1}{64} \left( u \ubar \right) \ln \left( \frac{\mu^2}{-\ubar p_1^2-u p_2^2} \right) \cdot \mathcal{R}_8 \quad 
\nn \\
&+\frac{1}{64} \left(3p^2+ u p^2 +4 u qp +4 u^2 qp\right) \ln \left( \frac{\mu^2}{-\ubar p_1^2-u p_2^2} \right)\cdot \mathcal{R}_9 \quad 
\nn \\
& \left. + \frac{1}{32} \left( u p^2 +3 u^2 p^2 +8 u^2 qp \right) \ln \left( \frac{\mu^2}{-\ubar p_1^2-u p_2^2} \right) \cdot  \mathcal{R}_{12} \right \rbrace 
\nn \\
&+\ldots \mbox{ .} \; \; \; \;
\end{align}

\item Diagram b:
\begin{align}
\label{Twist2}
T_b(q,p)=-\frac{e_u-e_d}{\pi^2} \qcond\int _0^1 \! \! du \; \Bigg[ 
\left(- \frac{1}{48}\left( \psi^{(A)}(u)+2\psi^{(V)}(u) \right)  \frac{1}{-\ubar p_1^1-u p_2^2} \right)\cdot \mathcal{R}_2 & \phantom{\Bigg]}
\nn\\ 
+\bigg( -\frac{1}{24}\left( \psi^{(A)}(u)+2\psi^{(V)}(u) \right)  \ln \left( \frac{\mu^2}{-\ubar p_1^1-u p_2^2}\right)\qquad\qquad\qquad&\phantom{\Bigg]}
\nn\\
-\frac{uqp}{24}\left( \psi^{(A)}(u)+2\psi^{(V)}(u) \right)  \frac{1}{-\ubar p_1^1-u p_2^2} \bigg)\cdot \mathcal{R}_3&\phantom{\Bigg]}
\nn\\ 
+\left( \frac{1}{24} \varphi(u) \ln \left( \frac{\mu^2}{-\ubar p_1^1-u p_2^2} \right)-\frac{\mathds{A}(u)} {32}\frac{1}{-\ubar p_1^1-u p_2^2}\right)\cdot \mathcal{R}_4 & \phantom{\Bigg]}
\nn\\ 
+\left( \frac{u \mathds{B}(u)} {16} \frac{1}{-\ubar p_1^1-u p_2^2} \right) \cdot \mathcal{R}_5 & \phantom{\Bigg]}
\nn \\ 
+\bigg(\frac{\varphi(u)}{6}(-\ubar p_1^2-up_2^2) \ln \left( \frac{\mu^2} {-\ubar p_1^1-u p_2^2} \right)\qquad\qquad\qquad& \phantom{\Bigg]}
\nn\\
 \quad\qquad\quad\qquad-\frac{\mathds{A}(u)+\mathds{B}(u)}{16} \ln \left( \frac{\mu^2} {-\ubar p_1^1-u p_2^2} \right) \bigg)\cdot \mathcal{R}_6 & \phantom{\Bigg]}
\nn \\ 
-\frac{\psi^{(A)} (u)+2\psi^{(V)}(u)} {24} \ln \left( \frac{\mu^2} {-\ubar p_1^1-u p_2^2} \right) \cdot \mathcal{R}_{7} & \phantom{\Bigg]}
\nn \\
-u\frac{\psi^{(A)}(u)+2\psi^{(V)}(u)} {48} \frac{1} {-\ubar p_1^1-u p_2^2}  \cdot \mathcal{R}_{8} & \phantom{\Bigg]}
\nn \\
+\bigg(-\frac{-5\psi^{(A)}(u)+2\psi^{(V)}(u)} {48} \ln \left( \frac{\mu^2} {-\ubar p_1^1-u p_2^2} \right)\qquad\qquad\qquad & \phantom{\Bigg]}
\nn\\
\quad\qquad\quad\qquad -uqp \frac{\psi^{(A)} (u)+2\psi^{(V)}(u)}{24}\frac{1} {-\ubar p_1^1-u p_2^2} \bigg) \cdot \mathcal{R}_{9} & \phantom{\Bigg]}
 \nn\\
+\frac{\mathds{B}(u)} {16} \frac{1}{-\ubar p_1^1-u p_2^2} \cdot \mathcal{R}_{10} & \phantom{\bigg]}
 \nn \\
+\left( \frac{u\varphi(u)}{6}  \ln \left( \frac{\mu^2}{-\ubar p_1^1-u p_2^2} \right)+\frac{u\mathds{A}(u)} {8}\frac{1}{-\ubar p_1^1-u p_2^2} \right)  \cdot \mathcal{R}_{11} & \phantom{\Bigg]}
\nn\\
+\bigg( -\frac{u\psi^{(A)}(u)+2u\psi^{(V)}(u)}{8} \ln \left( \frac{\mu^2}{-\ubar p_1^1-u p_2^2} \right)\qquad\qquad\qquad & \phantom{\Bigg]}
\nn\\
\quad\qquad\quad\qquad -\frac{u^2qp \left(\psi^{(A)}(u)+2\psi^{(V)}(u) \right)}{12} \frac{1}{-\ubar p_1^1-u p_2^2} \bigg)  \cdot \mathcal{R}_{12} \Bigg]
\nn\\
\qquad\qquad+\ldots \mbox{  .}\qquad\qquad\qquad \quad\qquad\quad\qquad\quad\qquad\quad\qquad\qquad\quad\qquad\quad\qquad&
\end{align}

\item Diagram c:
\begin{align}
\label{PhotonResult}
T_{c}(p,q)=&-\frac{(e_u-e_d)\qcond}{2 \pi^2}\int_0^1 \!\! du \int \!\!\Da  
\nn \\
&\times \Big \lbrace \left(-\frac{1}{4}\mathcal{S}_{\gamma}(\underline{\alpha})\frac{1}{-(p+\alpha_u q)^2}-\frac{1-2u}{4}\mathcal{T}_4^{\gamma}(\underline{\alpha}) \frac{1}{-(p+\alpha_u q)^2}\right)\cdot \mathcal{R}_4 +
\nn \\
&\phantom{\times \Big \lbrace} -\left(\frac{1}{2}\mathcal{S}_{\gamma}(\underline{\alpha})\frac{u \alpha_u}{-(p + \alpha_u q)^2}-\frac{1}{2}\mathcal{T}_4^{\gamma}(\underline{\alpha})\frac{u \alpha_u}{-(p + \alpha_u q)^2}\right) \cdot \mathcal{R}_5 \;\;
\nn \\
&\phantom{\times \Big \lbrace}+ \frac{\ubar}{2}\left(\mathcal{S}_{\gamma}(\underline{\alpha})+\mathcal{T}_4^{\gamma}(\underline{\alpha}) \right) \ln \left(\frac{\mu^2}{-(p+ \alpha_u q)^2} \right)\cdot \mathcal{R}_6 +
\nn \\
&\phantom{\times \Big \lbrace} +\frac{u}{2}\left(\mathcal{S}_{\gamma}(\underline{\alpha})+\mathcal{T}_4^{\gamma}(\underline{\alpha})\right)\frac{1}{-(p+ \alpha_u q)^2}\cdot \mathcal{R}_{10}\;\;
\nn \\
&\phantom{\times \Big \lbrace} -\left(\mathcal{S}_{\gamma}(\underline{\alpha})+\mathcal{T}_4^{\gamma}(\underline{\alpha})\right)\frac{\alpha_u}{-(p + \alpha_u q)^2} \cdot \mathcal{R}_{11} \Big \rbrace
\nn \\
&\phantom{\times \Big \lbrace} +\ldots \mbox{ .} \; \; \; \;
\end{align}

\item Diagram d:
\begin{equation}
\label{AllGluon}
T_{d}(p,q)=T_{G}^{(1)}(p,q)+T_{G}^{(2)}(p,q)+T_{G}^{(3)}(p,q)+T_{G}^{(4)}(p,q)+T_{d}^{(5)}(p,q) +\ldots \mbox{ .}
\end{equation}
With
\begin{align}
\label{gluon1}
T^{(1)}_{d}(p,q)= (e_u-e_d)\qcond \int_0^1 \!\!du \;\int \!\Da \; \mathcal{S}(\underline{\alpha}) \frac{1}{8 \pi^2} \bigg \lbrace - \frac{\ubar}{-(p+\alpha_uq)^2}\cdot \mathcal{R}_1&
\nn \\
 +(1-2u)\frac{\alpha_u}{-(p+\alpha_uq)^2}\cdot \mathcal{R}_5&
\nn \\
 +\left( \ln \left(\frac{\mu^2}{-(p+\alpha_uq)^2} \right)+(1-2u)\frac{2\alpha_u qp}{-(p+\alpha_uq)^2} \right) \cdot \mathcal{R}_6& 
\nn \\
 +\frac{u}{-(p+\alpha_uq)^2}\cdot \mathcal{R}_{10}& \bigg \rbrace
\end{align}

\begin{align}
\label{gluon2}
T^{(2)}_{d}(p,q)=(e_u-e_d) \qcond \int _0^1 \!\!du \int \!\Da\; \widetilde{\mathcal{S}}(\underline{\alpha})\;\;\bigg \lbrace \quad -\frac{\ubar}{8 \pi^2} \frac{1}{-(p+\alpha_u q)^2} \cdot \mathcal{R}_1 & 
\nn \\
+ \frac{1-2u}{8 \pi^2} \frac{1}{-(p+\alpha_u q)^2} \cdot \mathcal{R}_4 &
\nn \\
+ \frac{(1-2u)\alpha_u}{8 \pi^2} \frac{1}{-(p+\alpha_u q)^2} \cdot \mathcal{R}_5 &
\nn \\
+\left(-\frac{1}{8 \pi^2}\ln\left( \frac{\mu^2}{-(p+\alpha_u q)^2} \right)-\frac{\ubar \alpha_u qp}{4 \pi^2} \frac{1}{-(p+\alpha_u q)^2} \right)\cdot \mathcal{R}_6 &
\nn \\
+\frac{(1-2u)\alpha_u}{2 \pi^2} \frac{1}{-(p+\alpha_u q)^2}\cdot \mathcal{R}_{11}&\bigg \rbrace 
\end{align}

\begin{align}
\label{gluon3}
T^{(3)}_{d}(p,q)=(e_u-e_d)f_{3\gamma} \int_0^1\!\!du\;\int\!\!\Da\;\mathcal{V}(\underline{\alpha}) \bigg \lbrace\;\; \frac{\ubar}{4\pi^2}\frac{qp}{-(p+\alpha_uq)^2}\cdot\mathcal{R}_3& 
\nn \\
+\frac{u}{8\pi^2}\frac{1}{-(p+\alpha_uq)^2}\cdot \mathcal{R}_8 & 
\nn \\
+\left(-\frac{1}{4\pi^2}\ln\left(\frac{\mu^2}{-(p+\alpha_uq)^2}\right)+\frac{\ubar\alpha_u}{2\pi^2}\frac{qp}{-(p+\alpha_uq)^2} \right)\cdot \mathcal{R}_{12}& \bigg \rbrace
\end{align}

\begin{align}
\label{gluon4}
T^{(4)}_{d}(p,q)=(e_u-e_d) f_{3\gamma} \int_0^1 \! du \int \! \Da \;\mathcal{A}(\underline{\alpha}) \bigg \lbrace \;\frac{\ubar}{4 \pi^2} \frac{1}{-(p+\alpha_uq)^2} \cdot \mathcal{R}_{3}&
\nn \\
-\frac{u}{8 \pi^2}\frac{1}{-(p+\alpha_uq)^2}\cdot \mathcal{R}_8& 
\nn \\
-\frac{u}{2 \pi^2} \frac{qp}{-(p+\alpha_uq)^2} \cdot \mathcal{R}_9& 
\nn \\ 
+\left(-\frac{1}{4 \pi^2} \ln \left(\frac{\mu^2}{-(p+\alpha_uq)^2}\right)+\frac{(1-3u) \alpha_u}{2 \pi^2}\frac{qp}{-(p+\alpha_uq)^2} \right)\cdot \mathcal{R}_{12}& \;\bigg \rbrace
\end{align}

\begin{align}
\label{gluon5}
T^{(5)}_{d}(p,q)=-(e_u-e_d) \qcond \int_0^1 \! du  \! \;\;\;\;\;\; \qquad \qquad \qquad \qquad \qquad\qquad \qquad \quad \;\;\;&
\nn \\
\times\left \lbrace \int\Da \left(- \frac{u}{8 \pi^2} T_2(\underline{\alpha})+\frac{1-2u}{8\pi^2} T_3(\underline{\alpha}) -\frac{\ubar}{8 \pi^2} T_4(\underline{\alpha}) \right) \frac{1}{-(p+\alpha_uq)^2}\cdot \mathcal{R}_4  \;\! \right.& 
\nn \\
 +\left[\int \Da \left( \frac{\ubar \alpha_u}{8 \pi^2} T_2(\underline{\alpha})  -\frac{\ubar \alpha_u}{8 \pi^2} T_4(\underline{\alpha})   \right) \frac{1}{-(p+\alpha_uq)^2} \right.+\qquad \qquad &
\nn \\
 \left.+ \frac{\ubar}{2\pi^2} \left( \widetilde{I}^{(2)}-\widetilde{I}^{(4)} \right) \right] \cdot \mathcal{R}_5&
\nn \\
 +\int\Da\left[ \left( \frac{\ubar \alpha_u qp}{4\pi^4} T_2(\underline{\alpha})- \frac{\ubar \alpha_u qp}{4\pi^4} T_4(\underline{\alpha}) \right) \frac{1} {-(p+\alpha_uq)^2}\;\;\right. \qquad\qquad \qquad
\nn \\
 +\left( - \frac{1}{2\pi^2} T_1(\underline{\alpha})+\frac{5-6u}{8 \pi^2} T_2(\underline{\alpha})+\frac{1-2u}{4 \pi^2} T_3(\underline{\alpha}) -\frac{3-2u}{8 \pi^2}T_4(\underline{\alpha}) \right) \qquad  
\nn \\
 \times \left.\ln\left(\frac{\mu^2}{-(p+\alpha_uq)^2} \right)  \;\;\right]\cdot \mathcal{R}_6&
\nn \\
 +\int\Da  \left( \frac{u}{8 \pi^2} T_2(\underline{\alpha})- \frac{u}{8 \pi^2} T_4(\underline{\alpha}) \right)\cdot \mathcal{R}_{10}&
\nn \\
 +\left[\int\Da \left(-\frac{u \alpha_u}{2 \pi^2} T_2(\underline{\alpha}) + \frac{(1-2u)\alpha_u} {2 \pi^2} T_3(\underline{\alpha})-\frac{\ubar \alpha_u}{2 \pi^2}T_4(\underline{\alpha})\right)\frac{1}{-(p+\alpha_uq)^2} \right. \;\;\;\;
\nn \\
+\bigg (-\frac{1}{\pi^2} \left(\widetilde{I}^{(1)}+\widetilde{I}^{(4)} \right) +\frac{1-2u}{\pi^2} \left(\widetilde{I}^{(2)}+\widetilde{I}^{(3)} \right) \bigg) \bigg]\cdot \mathcal{R}_{11}& \bigg \rbrace \nn \\ 
\end{align}
where
\begin{align}
\int \Da&= \int_0^1 d\alpha_q \int_0^1 d\alpha_{\qbar} \int_0^1 d\alpha_g \; \delta(1-\alpha_q-\alpha_{\qbar}-\alpha_g) \\
\alpha_u&=\alpha_q+u\alpha_g \\
\widetilde{I}^{(i)}:=&\;  \ubar \int_0^1  \!d\alpha_q  \!\!\int_0^{\alpha_q}\!\!d\alpha_q'  \!\! \int_0^{1-\alpha_q'} \!\!d\alpha_{\qbar} \;\frac{T_i(\alpha_q',\alpha_{\qbar},1-\alpha_q'-\alpha_{\qbar})}{-(p+(u+\ubar \alpha_q-u \alpha_{\qbar})q)^2} 
\nn \\
& -u  \int_0^1\!\! d\alpha_q' \int_0^{1-\alpha_q'} \!\!d\alpha_{\qbar}\; \int_0^{\alpha_{\qbar}}\!\! d\alpha_{\qbar}' \;\frac{T_i(\alpha_q',\alpha_{\qbar}',1-\alpha_q'-\alpha_{\qbar}')}{-(p+(1-u \alpha_{\qbar}))^2} 
\nn \\
&+ u \int_0^{1}\!\! d\alpha_{q}' \; \int_0^{\alpha_q'}\!\!d\alpha_q'' \int_0^{1-\alpha_q''}\!\!d\alpha_{\qbar}'\;  \frac{T_i(\alpha_q'',\alpha_{\qbar}',1-\alpha_q''-\alpha_{\qbar}')}{-(p+(\ubar+u \alpha_q'))^2} \mbox{ .}
\end{align}
\end{itemize}

Here $\mu^2$ is an arbitrary scale and the dots denote polynomials in $p_1^2$ and $p_2^2$, that will vanish after a Borel transformation.
The $\widetilde{I}^{(i)}$ arise due to partial integration of terms proportional to $\frac{1}{qx}$.
The functions $\varphi(u)$, $\mathds{A}(u)$, $\mathds{B}(u)$, $\mathcal{S}(\underline{\alpha})$, $\widetilde{\mathcal{S}}(\underline{\alpha})$, $\mathcal{S}_{\gamma}(\underline{\alpha})$, $\psi^{(A)}(u)$, $\psi^{(V)}(u)$, $\mathcal{V}(\underline{\alpha})$, $\mathcal{A}(\underline{\alpha})$, $T_i(\underline{\alpha})$ and $\mathcal{T}_4^{\gamma}(\underline{\alpha})$ are defined in Appendix \ref{A}.

\section{The functions $\mathcal{I}_1$, $\mathcal{I}_2$ and $\mathcal{I}_3$}
\label{B2}
In this Section we have gathered the explicit expressions for the three functions $\mathcal{I}_1$, $\mathcal{I}_2$ and $\mathcal{I}_3$ that appear in Eqs.(\ref{G_1SR})--(\ref{G_1+G_2SR}).
\begin{align}
\mathcal{I}_1=& -\frac{v \mathds{A}(v)}{8} t \left( 1-e^{-S_0 / t} \right) 
\nn \\
& +\!\left[\phantom{-}\int_0^v\!\! d \alpha_q \int_0^{\overline{v}}\!\! d\alpha_{\qbar}\frac{v-\alpha_q}{2(1-\alpha_q-\alpha_{\qbar})^2} \mathcal{S}_{\gamma}(\alpha_q,\alpha_{\qbar},1-\alpha_q-\alpha_{\qbar}) \right.
\nn\\
&-\int_0^v d \alpha_q \int_0^{\overline{v}} d\alpha_{\qbar}\; v \frac{1-2v+\alpha_q-\alpha_{\qbar}}{2 (1-\alpha_q-\alpha_{\qbar})^2 } \widetilde{S}(\alpha_q,\alpha_{\qbar},1-\alpha_q-\alpha_{\qbar})
\nn  \\
&+ \int_0^v d \alpha_q \int _0 ^{\overline{v}} d\alpha_{\qbar} \; v\frac{1-2v+\alpha_q-\alpha_{\qbar}}{2(1-\alpha_q-\alpha_{\qbar})^2} T_3 (\alpha_q,\alpha_{\qbar},1-\alpha_q-\alpha_{\qbar}) 
\nn\\
& - \int_0^v d \alpha_q \int _0 ^{\overline{v}} d\alpha_{\qbar} \; \frac{v}{2(1-\alpha_q-\alpha_{\qbar})} T_4(\alpha_q,\alpha_{\qbar},1-\alpha_q-\alpha_{\qbar})
 \nn\\
&  - \int_0^v d \alpha_q \int _0 ^{\overline{v}} d\alpha_{\qbar} \; v\frac{v-\alpha_q} {2(1-\alpha_q-\alpha_{\qbar})^2} [T_2-T_4] (\alpha_q,\alpha_{\qbar},1-\alpha_q-\alpha_{\qbar})
\nn \\
& +\int_0^v d\alpha_q \int_0^{\alpha_q} d\alpha_{q}'\int_0^{\overline{v}} d\alpha_{\qbar}\; \frac{v-\alpha_q}{(1-\alpha_q-\alpha_{\qbar})^2}
\nn \\
& \quad \qquad \qquad \qquad \qquad  \qquad\times [-T_1+T_2+T_3-
T_4](\alpha_q',\alpha_{\qbar},1-\alpha_q'-\alpha_{\qbar})
\nn \\
&+\int_v^1 d\alpha_q \int_0^{\alpha_q} d\alpha_{q}'\int_{\overline{v}}^{1-\alpha_q'} d\alpha_{\qbar} \; \frac{v-\alpha_q}{(1-\alpha_q-\alpha_{\qbar})^2}
\nn \\
&\qquad \qquad \qquad \qquad \qquad  \qquad \times  [-T_1+T_2+T_3-
T_4](\alpha_q',\alpha_{\qbar},1-\alpha_q'-\alpha_{\qbar})
 \nn \\
&-\int_0^v d\alpha_q' \int_{\overline{v}}^{1-\alpha_q'} d \alpha_{\qbar} \int_0^{\alpha_{\qbar}} d \alpha_{\qbar}' \; \frac{ \overline{v}}{\alpha_{\qbar}^2}
\nn \\
&\quad \qquad \qquad \qquad  \qquad \qquad\times [-T_1+T_2+T_3-
T_4](\alpha_q',\alpha_{\qbar}',1-\alpha_q'-\alpha_{\qbar}')
\nn \\
&+\int_0^v d\alpha_q' \int_0^{\alpha_q'} d \alpha_q''\int_0^{1-\alpha_q''} d\alpha_{\qbar}' \; \frac{-\overline{v}}{(1-\alpha_q')^2}
\nn \\
& \quad \qquad \qquad \qquad \qquad  \qquad\times [-T_1+T_2+T_3-T_4](\alpha_q'',\alpha_{\qbar}',1-\alpha_q''-\alpha_{\qbar}')
\nn \\
&-2\int_0^v d\alpha_q \int_0^{\alpha_q} d\alpha_{q}'\int_0^{\overline{v}} d\alpha_{\qbar} \; \frac{(v-\alpha_q)(\overline{v}-\alpha_{\qbar})}{(1-\alpha_q-\alpha_{\qbar})^3}
\nn \\
& \quad \qquad \qquad \qquad \qquad  \qquad \times[T_2+T_3](\alpha_q',\alpha_{\qbar},1-\alpha_q'-\alpha_{\qbar})
\nn \\
&-2\int_v^1 d\alpha_q \int_0^{\alpha_q} d\alpha_{q}'\int_{\overline{v}}^{1-\alpha_q'} d\alpha_{\qbar} \; \frac{(v-\alpha_q)(\overline{v}-\alpha_{\qbar})}{(1-\alpha_q-\alpha_{\qbar})^3}
\nn \\
& \quad \qquad \qquad \qquad \qquad \qquad \times   [T_2+T_3](\alpha_q',\alpha_{\qbar},1-\alpha_q'-\alpha_{\qbar})
\nn \\
&+2\int_0^v d\alpha_q' \int_{\overline{v}}^{1-\alpha_q'} d \alpha_{\qbar} \int_0^{\alpha_{\qbar}} d \alpha_{\qbar}' \; \frac{ \overline{v}^2}{\alpha_{\qbar}^3}[T_2+T_3](\alpha_q',\alpha_{\qbar}',1-\alpha_q'-\alpha_{\qbar}')
\nn\\
&-2\int_0^v d\alpha_q' \int_0^{\alpha_q'} d \alpha_q''\int_0^{1-\alpha_q''} d\alpha_{\qbar}' \; \frac{-\overline{v}^2}{(1-\alpha_q')^3} [T_2+T_3](\alpha_q'',\alpha_{\qbar}',1-\alpha_q''-\alpha_{\qbar}') 
\nn\\
&\left.+\int_0^{v} d \alpha_q\int_0^{1-v} d\alpha_{\qbar} \; v\frac{1-v+\alpha_q-\alpha_{\qbar}}{2(1-\alpha_q-\alpha_{\qbar})^2} T_4^{\gamma}(\alpha_q,\alpha_{\qbar},1-\alpha_q-\alpha_{\qbar}) \right]
\nn\\
&\quad \qquad \qquad \qquad \qquad \qquad \qquad \qquad\times t \left( 1-e^{-S_0/t} \right) \\
\mathcal{I}_2=&-\frac{f_{3\gamma}}{8 \pi^2}\left( \frac{v}{3} \psi^{(V)}(v) + \frac{v}{6} \psi^{(A)}(v) \right) t \left( 1-e^{-S_0/t} \right)
\nn \\
& -\frac{f_{3\gamma}}{8\pi^2}\int_0^v d\alpha_q \int_0^{\overline{v}} d \alpha_{\qbar} \frac{v-\alpha_q}{(1-\alpha_q-\alpha_{\qbar})^2}
\nn \\
&\qquad\qquad\times\left[\mathcal{V}-\mathcal{A}\right](\alpha_q,\alpha_{\qbar},1-\alpha_q-\alpha_{\qbar}) t \left( 1-e^{-S_0/t} \right) 
\\
\mathcal{I}_3&=\int_0^v \!\!d \alpha_q \int_0^{\bar{v}} \!\!d \alpha_{\qbar}\; v\frac{v-\alpha_q}{(1-\alpha_{q}-\alpha_{\qbar})^2}
\nn \\
&\qquad\qquad\qquad\times \left[ \mathcal{S}_{\gamma}+\mathcal{T}^{\gamma}_4\right] (\alpha_q, \alpha_{\qbar}, 1-\alpha_q-\alpha_{\qbar}) t (1-e^{-S_0/t})
\nn \\
&\phantom{=}-\int_0^v \!\!d \alpha_q \int_0^{\bar{v}} \!\!d \alpha_{\qbar} \; v\frac{1-2v+\alpha_q-\alpha_{\qbar}}{2(1-\alpha_{q}-\alpha_{\qbar})^2}
\nn \\
&\qquad\qquad\qquad \times \left[\mathcal{S}+\tilde{\mathcal{S}}\right](\alpha_q, \alpha_{\qbar}, 1-\alpha_q-\alpha_{\qbar}) t (1-e^{-S_0/t})
\nn \\
&\phantom{=}-\int_0^v \!\!d \alpha_q \int_0^{\bar{v}} \!\!d \alpha_{\qbar} v\frac{\bar{v}-\alpha_{\qbar}}{2(1-\alpha_{q}-\alpha_{\qbar})^2} 
\nn\\
&\qquad\qquad\qquad \times \left[{T}_2-{T}_4\right](\alpha_q, \alpha_{\qbar}, 1-\alpha_q-\alpha_{\qbar}) t (1-e^{-S_0/t})
\nn \\
&\phantom{=}+2\int_0^v d\alpha_q \int_0^{\alpha_q} d\alpha_{q}'\int_0^{\overline{v}} d\alpha_{\qbar} \; \frac{(\overline{v}-\alpha_{\qbar})^2}{(1-\alpha_q-\alpha_{\qbar})^3}
\nn \\
& \qquad \qquad \qquad \times[T_2-T_4](\alpha_q',\alpha_{\qbar},1-\alpha_q'-\alpha_{\qbar})t (1-e^{-S_0/t})
\nn \\
&\phantom{=}+2\int_v^1 d\alpha_q \int_0^{v} d\alpha_{q}'\int_{\overline{v}}^{1-\alpha_q'} d\alpha_{\qbar} \; \frac{(\overline{v}-\alpha_{\qbar})^2}{(1-\alpha_q-\alpha_{\qbar})^3}
\nn \\
&\qquad \qquad \qquad \times   [T_2-T_4](\alpha_q',\alpha_{\qbar},1-\alpha_q'-\alpha_{\qbar})t (1-e^{-S_0/t})
 \nn \\
&\phantom{=}-2\int_0^v d\alpha_q' \int_{\overline{v}}^{1-\alpha_q'} d \alpha_{\qbar} \int_0^{\alpha_{\qbar}} d \alpha_{\qbar}' \; \frac{\overline{v}\left( \alpha_{\qbar}-\overline{v} \right)}{\alpha_{\qbar}^3}
\nn \\
& \qquad \qquad \qquad   \times [T_2-T_4](\alpha_q',\alpha_{\qbar}',1-\alpha_q'-\alpha_{\qbar}')t (1-e^{-S_0/t})
\nn\\
&\phantom{=}+2\int_0^v d\alpha_q' \int_0^{\alpha_q'} d \alpha_q''\int_0^{1-\alpha_q''} d\alpha_{\qbar}' \; \frac{-\overline{v}\left( v-\alpha_{q}'\right)}{(1-\alpha_q')^3}
\nn \\
& \qquad \qquad  \qquad \times [T_2-T_4](\alpha_q'',\alpha_{\qbar}',1-\alpha_q''-\alpha_{\qbar}') t (1-e^{-S_0/t})
\end{align}
The functions $\varphi(u)$, $\mathds{A}(u)$, $\mathds{B}(u)$, $\mathcal{S}(\underline{\alpha})$, $\widetilde{\mathcal{S}}(\underline{\alpha})$, $\mathcal{S}_{\gamma}(\underline{\alpha})$, $\psi^{(A)}(u)$, $\psi^{(V)}(u)$, $\mathcal{V}(\underline{\alpha})$, $\mathcal{A}(\underline{\alpha})$, $T_i(\underline{\alpha})$ and $\mathcal{T}_4^{\gamma}(\underline{\alpha})$ are again defined in Appendix \ref{A}.
\end{appendix}

\end{document}